%% file: PN24.tex
\documentclass[sigconf,authorversion,nonacm]{acmart}

\usepackage{booktabs} % For formal tables
\usepackage{float}
\usepackage{placeins}
\usepackage{svg}
\usepackage{tikz}
\usepackage{balance}
\usepackage[normalem]{ulem}

\newcommand\copyrighttextbottom{%
  \footnotesize Accepted version of M. Ge, B. Rossi, S. Chren, J. M. Blanco. Petri Nets for Smart Grids: The Story So Far. To appear in the proceedings of the 39th ACM/SIGAPP Symposium on Applied Computing (SAC '24), April 8--12, 2024, Avila, Spain. \url{https://doi.org/10.1145/3605098.3635989}}
  
\newcommand\copyrightnoticebottom{%
\begin{tikzpicture}[remember picture,overlay]
\node[anchor=south,yshift=14pt] at (current page.south) {\fbox{\parbox{\dimexpr1.0\textwidth-\fboxsep-\fboxrule\relax}{\copyrighttextbottom}}};
\end{tikzpicture}%
}

\begin{document}
\title{Petri Nets for Smart Grids: The Story So Far}

\author{Mouzhi Ge}
\affiliation{%
  \institution{Deggendorf Institute of Technology}
  \city{Deggendorf}
  \country{Germany}
}
\email{mouzhi.ge@th-deg.de}

\author{Bruno Rossi}
\affiliation{%
  \institution{Masaryk University}
  \city{Brno} 
  \country{Czech Republic}
}
\email{brossi@mail.muni.cz}

\author{Stanislav Chren}
\affiliation{%
  \institution{Aalto University}
  %\streetaddress{P.O. Box 1212}
  \city{Espoo} 
  \country{Finland}
}
\email{stanislav.chren@aalto.fi}

\author{Jos\'e Miguel Blanco}
\affiliation{%
  \institution{Universidad Polit\'ecnica de Madrid}
  \city{Madrid}
  \country{Spain}
}
\email{josemiguel.blanco@upm.es}

\renewcommand{\shortauthors}{}

\begin{abstract}
Since the energy domain is in a transformative shift towards sustainability, the integration of new technologies and smart systems into traditional power grids has emerged. As an effective approach, Petri Nets (PN) have been applied to model and analyze the complex dynamics in Smart Grid (SG) environments. However, we are currently missing an overview of types of PNs applied to different areas and problems related to SGs. Therefore, this paper proposes four fundamental research questions related to the application areas of PNs in SGs, PNs types, aspects modelled by PNs in the identified areas, and the validation methods in the evaluation. The answers to the research questions are derived from a comprehensive and interdisciplinary literature analysis. The results capture a valuable overview of PNs applications in the global energy landscape and can offer indications for future research directions. 

\end{abstract}

\maketitle
\copyrightnoticebottom

\section{Introduction} 
Smart Grids (SG) represent one of the most essential Cyber-physical Critical Infrastructures~\cite{AhmadGAA23}. SGs were born in close connection to providing consumers with a more sustainable and secure electricity supply supported by renewable energy sources~\cite{yu2011new}. However, the SG represents the confluence of many \textit{"cyber"} aspects (e.g., computing algorithms, communication, and control software) together with physical parts (e.g., sensors, smart meters, physical devices)~\cite{farhangi2009path}. Such confluence represents the complexity of the whole domain, and in turn leads to crucial challenges for modelling and representing the whole SG infrastructure and its ongoing processes.

Petri Nets (PN) constitute a relatively old formalism that has found wide adoption in many domains to solve various problems connected to modelling components and processes~\cite{JengD93}. As expected, also in the context of SGs, they found wide adoption - for example, to model security constraints (e.g., for analyzing and modelling cyber-physical attacks and countermeasures \cite{liu2017modeling}) or to model reliability aspects (e.g., for faults localization and recovery  \citep{Sreerama2017}). Still, we need a broad overview of all the aspects modelled by means of PNs in the SGs context, the specific type of PN applied, and the main research results. Such an overview can allow researchers to understand the current challenges better and delineate future research directions.

PNs have been integrated and used in SGs in recent research. For example, PNs can model and analyze the behavior of SGs components, such as power generation units, energy storage systems, and distribution networks \cite{ref:Contreras2023}. Also, PNs can be utilized for system simulation and performance analysis, allowing grid operators to assess the impact of different scenarios and strategies \cite{Zeineb2016}. Furthermore, PNs can help to detect faults and disturbances in SGs by modelling the behavior of the grid components and their interactions \cite{mouzhireadinprogress}. Integrating PNs analysis in SGs provides an effective way to model, analyze, and control the electrical grids. SGs can also be effectively simulated, optimized, and managed by PNs. 

Considering the emerging research of PNs in SGs, the main contribution of the paper is to review and map the current research on the usage of PNs in the context of SGs. We cover several aspects related to the application of PNs in this domain: 
\begin{enumerate}
    \item The areas in which Petri Nets have been applied in the context of Smart Grids (e.g., \textit{SGs security});
    \item The types of Petri Nets that are mostly applied in the Smart Grids domain (e.g., \textit{Stochastic PNs});
    \item  Aspects that are modeled by Petri Nets in the identified areas (e.g., \textit{modelling of cyber-attacks});
    \item Validations that have been used to evaluate the Petri Nets (e.g., \textit{tools or datasets adopted for the analysis});
\end{enumerate}

The paper is structured as follows: In Section \ref{sec:background} we provide the background about PNs. We also review the concept of SGs with the focus at the level of the overall architecture and different layers of the infrastructure. In Section \ref{sec:rel-works} we provide results from related works in terms of previous surveys and research about the adoption of PNs in the SGs context. In Section \ref{sec:research-methods} we provide the method adopted for the review and the description of the whole review process. In Section \ref{sec:results} we provide a mapping of research areas related to PNs and SGs in the context of categories extracted from the papers (\textit{networking}, \textit{management}, \textit{security}, \textit{reliability}) with answers to the main RQs of the paper. In Section \ref{sec:conclusion} we conclude the paper summarising the main results.

\section{Background} 
\label{sec:background}
 \subsection{Petri Nets} 

PNs are a modelling formalism developed by C. A. Petri in 1962 in his thesis entitled \emph{Communication with Automata}. Since then, the PNs have been in used to solve various theoretical and application problems in numerous domains  \cite{JengD93}. A PN is a bipartite directed graph where nodes belong to one of the disjoint sets - places and transitions. The edges are represented by directed arcs that connect places to transitions (input arcs) or vice versa (output arcs). In a PN, tokens can be associated with places which results in a marked PN. The vector of the number of tokens in each place is called marking of the PN. 
A PN can be defined formally as a 4-tuple $(P, T, F, M_0)$, where $(P, T, F)$ is a net and $M_0$ represents the initial markings \cite{Rozenberg1998}. While $P$ is a finite set of places that represents the states or conditions of a system, $T$ is a finite set of transitions, and it represents actions taken or events occurred in the system. Furthermore, $F  \subseteq  (P \times T) \cup (T \times P)$ is the set of arcs, also known as flow relation. $F$ defines the relations between places and transitions. $M_0: P \to \mathbb{N}_0$, where $\mathbb{N}_0$ is the vector of initial state that is described by marking such as the initial distribution of tokens in the places.

PNs provide a graphical representation of a system's state transitions \cite{JengD93}. The basic components of a PN are places, transitions, and arcs, where places represent states or conditions of the system, transitions represent events or actions, and arcs define the relationships between places and transitions \cite{Ireaditandindeeditisclassic}. Further, tokens are used to represent the state of a place. 
The interaction in the PN is achieved via firing of the enabled transitions. A transition is considered enabled in the given marking if each input place of the transition contains at least one token.
Based on those components, PNs can be used to analyze complex systems with multiple interacting components. Particularly, PNs are useful when we model and analyze systems that have parallelism, synchronization, and resource allocation as key elements \cite{mouzhianotheronereadinprogress}.

One of the advantages of PNs is to accurately capture the dynamic behavior of systems. They can model concurrent processes and represent multiple activities simultaneously. Given the intuitive visualization feature, PNs can facilitate the system design, analysis and communications among stakeholders. PNs have been used to analyze various problems, such as reachability analysis, deadlock detection, liveness analysis, and performance evaluation \cite{HujsaD18}. 
With the development of their usage, PNs have been evolved from standard PNs to several extensions, which are designed to cater specific requirements and extend modelling capabilities. Some widely used types of PNs are as follows. 
\begin{enumerate}
    \item Standard Petri Nets: Also known as Place/Transition (P/T) Nets, Standard Petri Nets are the basic form of PNs, and it is mainly used for modeling concurrent and asynchronous systems.
    \item Stochastic Petri Nets (SPN): SPNs introduce probabilistic behavior to standard PNs \cite{SPN}. They connect probabilities with transitions. Thus, it enables users to model non-deterministic and uncertain systems. SPNs are used for performance analysis, reliability assessment, and system optimization. 
    \item  Timed Petri Nets (TPN): TPNs include time into the PN modelling \cite{TPN}. They use the concept of time delays associated with transitions. Thus, users can represent time-dependent behaviors, synchronization, and performance analysis. TPNs are widely used in real-time systems and protocols.
    \item Colored Petri Nets (CPN): CPNs extend the ability of PNs by associating tokens with specific colors or attributes \cite{CPN}. These attributes can represent various properties, such as resource types, values, and conditions. Thus, CPNs can be used for detailed modeling and analysis, particularly, they are useful for modeling complex systems with data-related behavior.
    \item Fuzzy Petri Nets (FPN): FPNs combine the concepts of PNs with fuzzy logic, which provide a modelling framework to handle imprecision, uncertainty, and vagueness in system behavior \cite{FPN}. One of the merits of FPNs is that they are able to capture and reason about linguistic variables and fuzzy rules. Thus, FPNs have been used in more realistic and flexible modeling of complex systems in uncertain environments such as robotics, process control, and AI systems.
    \item Hybrid Petri Nets (HPN), HPNs are the extension of PNs that combine the modelling of discrete and continuous events \cite{HPN}. Since in real-world systems continuous variables come along together discrete events, HPNs are then developed to address this type of modelling by modeling both discrete and continuous event within a unified framework. HPNs have been used in different domains such as manufacturing systems, biological systems, and cyber-physical systems. 
\end{enumerate}

Each type of PN offers unique features and advantages, which can be used to model and analyze various system behaviors such as concurrency, timing, resource allocation, probabilistic behavior, and interactions. The selection of PN type depends on the specific requirements and characteristics of the system.

\subsection{Smart Grids} 
The requirements for electricity continue to grow exponentially due to population growth, urbanization, and the increasing reliance on technology in our daily lives. Traditional power grids face significant challenges to meet the growing demand, such as old infrastructure, inefficient energy distribution, and limited integration of renewable energy sources. To address these challenges, the concept of SGs has emerged as a transformative solution to revolutionize generating, distributing, and consuming electricity~\cite{AhmadGAA23}.

A SG is an advanced electrical grid system that leverages new technologies, communication networks, and intelligent devices to enable a more efficient, reliable, and sustainable electric power infrastructure~\cite{SuhaimyRAAH22}. It represents a significant advancement from the conventional and one-way power flow model by transforming it into a dynamic and interactive network that seamlessly integrates various energy sources, such as solar and wind while enabling users with flexible control over their energy consumption~\cite{MedinaAR23}.

SGs enable grid operators to manage electricity generation and distribution in an intelligent manner; for example, in SG networks, the sensors, meters, and automation technologies provide data on energy consumption patterns, voltage levels, and equipment performance~\cite{SpringmannBM22}. Thus, it enables us to develop proactive grid management, timely fault detection, and optimized load balancing. Also, SGs can minimize power outages, reduce downtime, and improve the overall resilience of the electrical grid. Since SGs rely on communication infrastructure that enables real-time information exchange between different components of the electrical grid, such as power generation units, substations, distribution networks, and users, it allows for improved monitoring and control, and it also leads to operational efficiency and better reliability~\cite{GongZYLLY21}. 

\begin{figure}[htb]
  \begin{center}
    \includegraphics[width=0.9\linewidth]{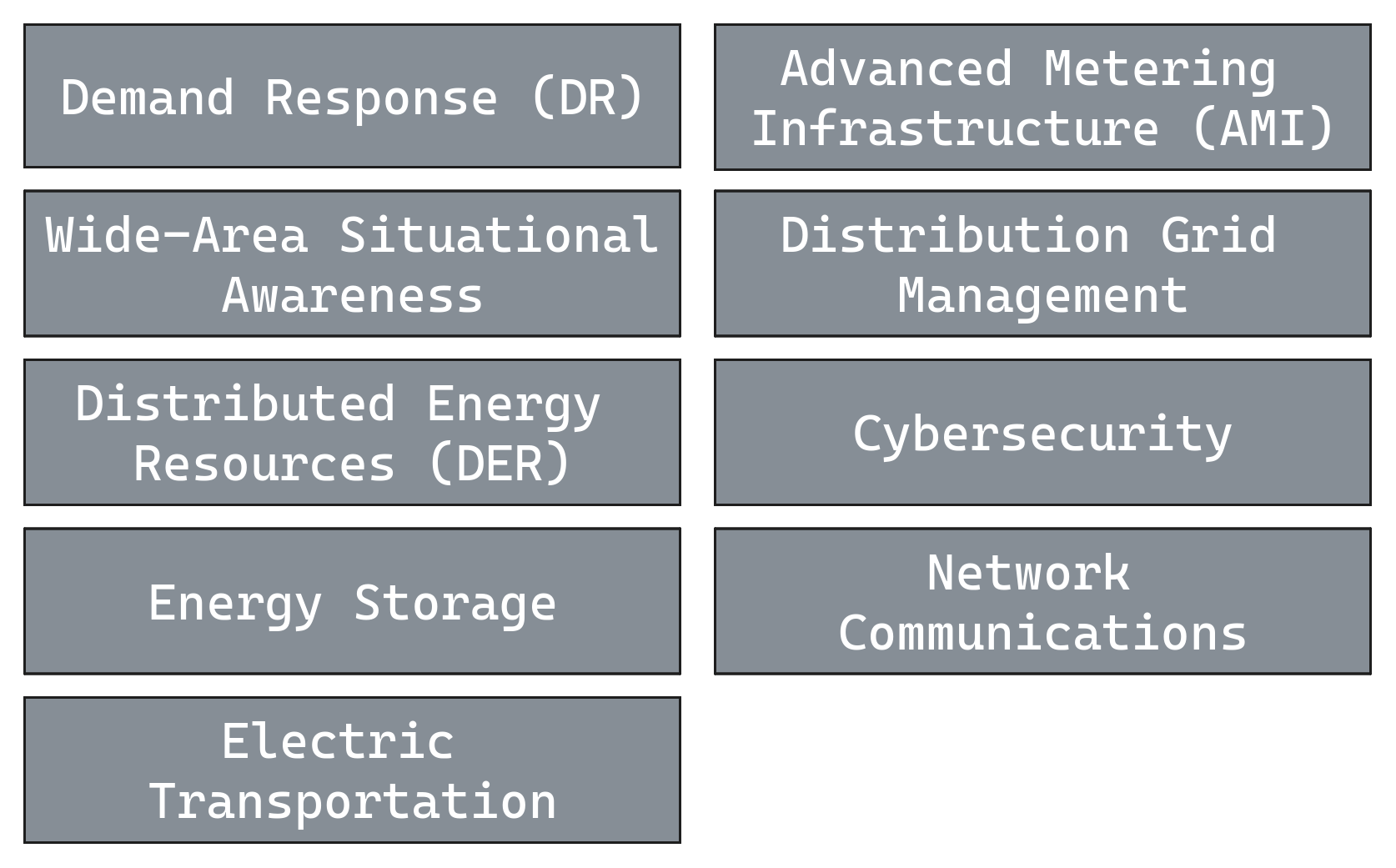}
  \end{center}
  \caption{SG research areas according to Cintuglu et al.~\cite{ref:cintuglu2017survey}}
  \label{fig:SG-research-areas}
\end{figure}

From the user perspective, SGs enable consumers to actively participate in managing their energy usage and costs. Through smart meters and home energy management systems, users may have access to real-time data about their electricity consumption~\cite{ShateriMPL20}, thus, it enables them to make effective decisions, for example, adjusting their energy usage during peak demand periods and potentially saving money.

SGs have shifted the way we manage and use electricity: They increase the grid reliability and optimize energy distribution, while they also facilitate the development of microgrids and virtual power plants~\cite{AhmadSAA23}, which can connect the collective power of distributed and renewable energy resources to ensure a reliable and sustainable energy supply.

In \cite{ref:cintuglu2017survey} there are nine areas defined for SGs research (Fig.\ref{fig:SG-research-areas}).

\begin{enumerate}
    \item Demand Response (DR) is related to the real-time balancing power demand and supply.
    \item Wide-Area Situational Awareness deals with the prevention and response to events to avoid any SG disruption.
    \item Distributed Energy Resources (DER) are related to the energy generation to supply power capacity. 
    \item Energy Storage, that is  converting electrical energy from the power network into modalities for storage.
    \item Electric Transportation, deals with all the aspects of the integration of electric vehicles with the grid.
    \item Advanced Metering Infrastructure (AMI) is the technological connection between consumers and distribution system operators (DSO).
    \item Distribution Grid Management deals with all the power distribution operations and optimizations.
    \item Cybersecurity relates to all the activities for the detection, prevention, and hampering of security vulnerabilities in SGs.
    \item Network Communication deals with the communications in SGs over wireless as well as wired networks.
\end{enumerate}

\section{PN and SG: what we know so far} 
\label{sec:rel-works}

The question about the previous systematization of knowledge that deals with SGs has to be answered in two different ways. The first one relates to data analytics in SGs. Previous reviews cover different topics, such as data analytics \cite{ref:alahakoon2016futuresurvey, ref:bhardwaj2017smartgridsanalyticssystematic, ref:pawar2017data-analytics-brief-review, ref:rossi2020}, big data \cite{ref:diamantoulakis2015big, ref:ge2018big, ref:hu2016bigdatachallenges, ref:tu2017bigdatareview}, load management strategies \cite{ref:benetti2016slr}, or the communication infrastructure of the grid \cite{ref:yan2013communicationsurvey}. Nevertheless, none of these previous efforts have PNs as a central focus. If PNs are mentioned, it is only because they play a role in the models used. This means that none of the aforementioned reviews is specifically targeted at understanding the application of PNs in the SGs infrastructure.

The second topic we have to address is related to the reviews that focus on applying PNs in the context of SGs. However, we have yet to find any previously rigorously conducted review looking at the specificity of the application of PNs in this domain. In particular, the existing reviews address the usage of PNs for specific aspects of SGs (e.g., \cite{Calderaro2011} focuses on failure identification, \cite{Fendri2020a} focuses on isolated microgrids, \cite{wang2015a} focuses on fault diagnosis, and \cite{Zeineb2016} focuses on performance analysis). Not even the most recent ones (e.g., \cite{ref:Contreras2023} focusing on control models) tackle the full usage of PNs.

By observing the works above, we conclude that the goal of providing the first in-depth overview of PNs usage within the domain is not out of line. It will offer insight into the most applied PNs types, problems solved, results, and tools used. The literature review allows a mapping of the overall research area and, therefore, does not restrict the review to one aspect (e.g., \textit{AMI}). The review will cover PNs as applied to the whole SG infrastructure, allowing for a less fine-grained discussion of sub-areas.

\section{Research Method} 
\label{sec:research-methods}
We run a scoping literature review~\cite{munn2018systematic} that can be adopted to define the focus of research in a given field.
The scoping review is based on three main phases. In phase~I, we define the Research Questions (RQs). Based on these, the RQs, we define a set of relevant keyword queries to four digital repositories. After obtaining the initial set of papers, we remove duplicates, and identify relevant articles by full papers read. Outgoing/incoming citation for the most cited articles in each category are used to improve the quality of the results. In phase~II, the categories of articles are extracted by means of open coding by defining macro-categories. We based the categorization partially on the research of \cite{ref:cintuglu2017survey}, but having four macro categories rather than nine categories. In phase~III, we answer the RQs by extracting and aggregating the information from the research articles.

 \subsection{Research Questions (RQs)}
The aim of this study is to obtain an overview of the application of PNs in the context of SGs. As discussed in ~\cite{ref:cintuglu2017survey}, there are multiple research areas that need to be considered. Thus, we targeted the RQs to the identification of area of research of application of PNs, the models covered, the techniques used, and the analysis settings when applied to case studies. To drive the review process, we propose the following RQs. 

\begin{enumerate}
\item[{RQ}1.] \textit{What are areas in the smart grid domain in which the Petri Nets are applied?} 
\item[{RQ}2.] \textit{Which types of Petri Nets are applied in the smart grid domain?} 
\item[{RQ}3.] \textit{What is modeled by Petri Nets in the identified areas?} 
\item[{RQ}4.] \textit{What type of evaluation has been carried out for the use of the Petri Nets?} 
\end{enumerate}

The rationales for proposing the RQs above are to investigate in which areas of the SGs infrastructure PNs have been more applied over the years~(RQ1), which types of PNs are mostly adopted~(RQ2), what are PNs used to model in the related research~(RQ3), and to obtain an overview of the details of case studies in which PNs have been applied~(RQ4). Finally, it is important to understand the implications of the PNs application in each area by aggregating the information from all the RQs.

\subsection{Review Process}
The search process is based on the definition of keywords used in the RQs, and on running the queries on a set of digital repositories identified in the process. Overall, there is a close connection between the RQs and the queries used for the review process. This process is fine-tuned based on the initial search queries to look for consistent results from the different digital repositories and was complemented with additional steps, such as looking at in-out-going citations once an initial set of papers is gathered. These additional steps allow to increase the quantity of research reviewed, as ensuring completeness of the reviewed papers is challenging due to the differences in digital repositories and the amount of research published yearly.

We selected four digital repositories that we consider relevant for the search type. We excluded general repositories such as Google Scholar, as they primarily collect articles from other repositories, which would have led to many duplications. We restricted the review to the following major repositories: IEEExplore, ACM Digital Library, Elsevier ScienceDirect, and SpringerLink. For each repository, we run the queries on abstract, keywords, and title. The only exception is the SpringerLink repository, that allowed only a full-text search. The key query terms are “Petri Net” AND “Smart Grid”, “Petri Net” AND “Power Grid”, as well as “Petri Net” AND “Energy”. 

In the search results, we only consider the papers that discuss the application of PNs regarding the SG infrastructure, and the papers should provide empirical results/findings for the application of PNs in SGs. Additionally, the papers should be published in English and either conferences, workshops, or academic journals. We exclude papers not dealing with the SG domain, with areas defined as in~\cite{ref:cintuglu2017survey}. Also, editorials, reports, and keynotes are not considered.

\section{Results}
\label{sec:results}

%-----------------
\begin{enumerate}
\item[\textbf{{RQ}1}.] \textit{What are areas in the smart grid domain in which the Petri Nets are applied?}
\end{enumerate}

Based on our review, we have found four important areas where PNs are applied in SGs: \textit{management}, \textit{reliability}, \textit{networking} and \textit{security}. The scope of the four identified areas are derived as follows: PNs provide an effective approach for modeling in the management area of SGs. They can be used to model and simulate different operational processes such as resource allocation or energy scheduling. One of the important goals of using PNs in SGs is to achieve reliability. The reliable operation of the SG is critical to assure continuous power supply and minimizing downtime, where PNs can be used to model and analyze the grid's reliability aspects such as fault detection and problem diagnosis. Apart from reliability, efficient communication among various components is important in SGs. To this end, PNs can be used to analyze the network status, and traffic patterns in the SGs. With the increase of digital communications in SGs, security has become a significant challenge in SGs. Thus, PNs can be used to conduct intrusion detection, thread detection as well as vulnerability assessment. If we compare the categorization used in this paper with the one in~\cite{ref:cintuglu2017survey}, we have the same \textit{networking} (\textit{network communication} in~\cite{ref:cintuglu2017survey}) and \textit{security} (\textit{cybersecurity} in~\cite{ref:cintuglu2017survey}) categories, while \textit{management} and \textit{reliability} are more cross-cutting categories that we preferred over more fine-grained categories to ease the scoping review.  
\subsubsection*{Management}
The area of management of SGs encompasses several aspects related to the organization and running of the SGs infrastructure. Usage of PNs in the area is related to control automation, power balancing, load control in microgrids~\cite{Chamorro2012a,Chamorro2012}, power balance of production/consumption of hybrid AC/DC microgrids~\cite{Qachchachi2016,DeBenedictis2018}, adaptive autonomy for utility management automation and to take into account experts opinions and environmental condition~\cite{Zamani2010,Fereidunian2011},integration of Electrical Vehicles for balancing power consumption/production~\cite{Kaur2016,Tolosana-Calasanz2013}, strategies for better structuring the SGs Infrastructure~\cite{Zamani2011,Zeineb2016}.

\subsubsection*{Reliability} 
The application of PNs in the area of reliability for SGs provides valuable insights into the functioning of the infrastructure in SGs. By modeling and simulating different aspects such as failure detection, fault diagnosis, dependability analysis, resilience, and load balancing, PNs can be used to support the development of reliable and efficient SGs. \cite{Liu2010,Calderaro2011,Wang2014,Xu2019} proposed fault diagnosis models in the SG network by PNs. \cite{Zeng2011,Zeng2012} introduced a model for dependability analysis of control centre networks using GSPN. To show the cost effectiveness of the solution, \cite{Sreerama2017,Kiaei2020} proposed PNs models for fault localization in SGs.

\subsubsection*{Networking} Since SGs extensively rely on a complex network of communication technologies to monitor, control, routing, and coordinate various grid components, PNs can be used to model and analyze the behavior of the communication networks. Also, PNs can provide detailed analysis regarding the network performance and communications. An effective and efficient communication is crucial in the distributed control systems in the SGs. Therefore, \cite{JAMRO201577,Buccafurri2019} investigate how to develop efficient and resilient communication for distributed control systems in SGs. Further, \cite{Buccafurri2019} proposed a routing algorithm that aims at individuating different communication paths in a smart distribution network. \cite{Mahendran14} is to address how to analyze the performance of networks in SGs. 

\subsubsection*{Security} The question about the value of security regarding SGs is out of doubt, since they constitute a critical infrastructure whose correct operation is considered as crucial due to the dependability that some of the other critical infrastructures. Thus, in order to prevent possible cascading effects, the security of the grid is a topic of maximum importance that is also addressed by the use of PNs. In this regard, the usage of PNs is mostly devoted to the modelling of different attacks. These attacks can be cyber in nature, as topology attacks such as in \cite{li2018reliability,Tare2016} addressing the delay of messages, or \cite{zaitsev2016} targeting the disguised traffic attacks. Or they can be of cyber-physical attacks as in \cite{Chen2011} focusing on the weaknesses of smart meters, or  in \cite{xu2018petri} targeting the modelling of the impact. The last kind of attacks that are modelled by PNs are insider threats, as it can be seen in \cite{Li2019a}. A final topic that is also covered by the usage of PNs is the modelling of the security system as in \cite{RafaeldaSilva2017}.

\begin{table*}[htb]
\caption{Distribution of articles by SG Research Area }
\centering
\footnotesize
\begin{tabular}{|p{1.3cm}|p{2cm}|p{3cm}|p{5cm}|p{4.5cm}|}
\hline
\textbf{Area} &
  \textbf{Papers} &
  \textbf{Type of PN} &
  \textbf{Aspects modelled} &
  \textbf{Evaluation} \\ \hline
Management &
  \cite{Bouazza2019}, \cite{Kim2019}, \cite{Kaur2016}, \cite{Gentile2014}, \cite{Congqi2013}, \cite{Chamorro2012a}, \cite{Zamani2011}, \cite{Munoz2010}, \cite{Zamani2010}, \cite{Lu2010}, \cite{Zeineb2016} &
  \textbf{Standard:} \cite{Bouazza2019}, \cite{Kim2019}, \cite{Chamorro2012a}, \cite{Zamani2011}, \cite{Lu2010} \textbf{Coloured:} \cite{Kaur2016}, \cite{Congqi2013}, \cite{Munoz2010}, \cite{Zeineb2016} \textbf{Stochastic:} \cite{Gentile2014} \textbf{Timed:} \cite{Zamani2010} &
  \textbf{Temperature control in smart buildings:} \cite{Bouazza2019}, \textbf{Bus voltage control:} \cite{Kim2019}, \textbf{Reduction of power fluctuations:} \cite{Kaur2016}, \textbf{Maximise acquired energy in smart homes:} \cite{Gentile2014}, \textbf{Stability of control in microgrids:} \cite{Congqi2013}, \textbf{Load Control in Microgrids:} \cite{Chamorro2012a}, \textbf{Optimization of IT infrastructure for SG:} \cite{Zamani2011}, \textbf{Power distribution network:} \cite{Munoz2010}, \textbf{Adaptive autonomy:} \cite{Zamani2010}, \textbf{Energy management of power station:} \cite{Lu2010}, \textbf{Energy management of smart grids:}  \cite{Zeineb2016} &
  \textbf{Scenario} - artificial: \cite{Gentile2014}, \cite{Zamani2010}, \cite{Zamani2011}, \cite{Zeineb2016} ; realistic: \cite{Chamorro2012a}, \cite{Munoz2010}, \cite{Congqi2013}, \cite{Kaur2016}, \cite{Kim2019}, \cite{Bouazza2019}; real: \cite{Lu2010}. \textbf{Data} - artificial: \cite{Gentile2014}, \cite{Zamani2010}, \cite{Zamani2011}, \cite{Zeineb2016} ; realistic: \cite{Chamorro2012a}, \cite{Munoz2010}, \cite{Congqi2013}, \cite{Kaur2016}, \cite{Kim2019} ; real: \cite{Bouazza2019}, \cite{Lu2010}. \textbf{Evaluation} - demonstration-simple: \cite{Gentile2014}, \cite{Zamani2010}, \cite{Lu2010}, \cite{Zamani2011}, \cite{Zeineb2016} ; demonstration complex: \cite{Chamorro2012a}, \cite{Munoz2010}, \cite{Bouazza2019}, \cite{Congqi2013}, \cite{Kim2019} ; comparison: \cite{Kaur2016} \\ \hline
Reliability &
  \cite{Liu2010}, \cite{Calderaro2011}, \cite{Zeng2012}, \cite{Diekhake2013}, \cite{Wang2014}, \cite{Wang2014a}, \cite{wang2015a}, \cite{Ghasemieh2015}, \cite{Chen2015}, \cite{Panchal2016}, \cite{Huels2016}, \cite{Marrone2016}, \cite{Matos2016}, \cite{Morris2017}, \cite{Mahdi2017}, \cite{Sreerama2017}, \cite{Jiang2018}, \cite{Xu2019}, \cite{Kiaei2020} &
  \textbf{Standard:} \cite{Saki2011}, \cite{Liu2010}, \cite{Calderaro2011}, \cite{wang2015a}, \cite{Panchal2016}, \cite{Sreerama2017}, \cite{Jiang2018} \textbf{Stochastic:} \cite{Zeng2011}, \cite{Zeng2012}, \cite{Wang2014a}, \cite{Morris2017}, \cite{Mahdi2017} \textbf{Fluid Stochastic:} \cite{Huels2016},  \cite{Marrone2016} \textbf{Hybrid:} \cite{Ghasemieh2015}, \cite{Matos2016} \textbf{Fuzzy:} \cite{Wang2014}, \cite{Chen2015}, \cite{Xu2019}, \cite{Xu2019}, \cite{Kiaei2020} \textbf{Causal:} \cite{Diekhake2013} &
  \textbf{Monitoring and failure diagnostics:} \cite{Saki2011}, \textbf{Dependability  analysis:} \cite{Zeng2011}, \cite{Zeng2012},  \textbf{Fault identification and diagnostics:} \cite{Liu2010}, \cite{Calderaro2011}, \cite{Wang2014}, \cite{Jiang2018}, \cite{Sreerama2017}, \cite{Xu2019}, \cite{Kiaei2020}, \textbf{Monitoring:} \cite{Diekhake2013}, \textbf{Detection of nontechnical losses, outages, illegal and fault events:} \cite{Chen2015}, \textbf{Reliability and risk analysis:} \cite{Panchal2016}, \textbf{Resilience:} \cite{Huels2016}, \textbf{Control logic adaptation:} \cite{Marrone2016}, \textbf{Fault tolerance, fault recovery:} \cite{Matos2016}, \textbf{Availability and resiliency analysis:} \cite{Morris2017}, \textbf{Reliability and availability analysis:} \cite{Mahdi2017} &
  \textbf{Scenario} - artificial: \cite{Liu2010}, \cite{Zeng2012}, \cite{Wang2014}, \cite{Panchal2016}, \cite{Matos2016}, \cite{Mahdi2017}, \cite{Sreerama2017}, \cite{Jiang2018}; realistic: \cite{wang2015a}, \cite{Chen2015}, \cite{Huels2016}, \cite{Morris2017}, \cite{Xu2019}, \cite{Kiaei2020}; real: \cite{Ghasemieh2015},  \cite{Calderaro2011}, \cite{Diekhake2013}. 
\textbf{Data} - artificial:  \cite{Liu2010}, \cite{Zeng2012}, \cite{wang2015a}, \cite{Chen2015}, \cite{Matos2016}, \cite{Mahdi2017}, \cite{Jiang2018},  \cite{Kiaei2020}; realistic: \cite{Calderaro2011}, \cite{Wang2014},  \cite{Panchal2016}, \cite{Morris2017}, \cite{Sreerama2017} ; real: \cite{Diekhake2013}, \cite{Huels2016}, \cite{Xu2019}, \cite{Ghasemieh2015}. 
\textbf{Evaluation} - demonstration-simple:  \cite{Liu2010}, \cite{Zeng2012}, \cite{Wang2014}, \cite{wang2015a}, \cite{Panchal2016}, \cite{Matos2016}, \cite{Mahdi2017}, \cite{Jiang2018}, \cite{Kiaei2020}; demonstration complex:  \cite{Calderaro2011}, \cite{Ghasemieh2015}, \cite{Chen2015}, \cite{Huels2016}, \cite{Morris2017}, \cite{Sreerama2017}, \cite{Xu2019}; comparison: \cite{Ghasemieh2015}\\ \hline
Networking &
  \cite{Machado2018}, \cite{JAMRO201577}, \cite{Lei13}, \cite{Peng2019}, \cite{Zhang2018a}, \cite{Xiang2014}, \cite{Fendri2019} &
  \textbf{Stochastic:} \cite{Lei13}, \cite{Peng2019}, \cite{Xiang2014} \textbf{Coloured:} \cite{Machado2018},  \cite{JAMRO201577} \textbf{Timed Fuzzy:} \cite{Zhang2018a} \textbf{Timed Hybrid:} \cite{Fendri2019} &
  \textbf{Distributed energy resource:} \cite{Machado2018}, \textbf{Communication in control systems:} \cite{JAMRO201577}, \textbf{Device-to-Device communications:} \cite{Lei13}, \textbf{Network scheduling strategy:} \cite{Peng2019}, \textbf{real-time and multi-source sensing:} \cite{Zhang2018a}, \textbf{Resource optimization and communication:} \cite{Xiang2014}, \textbf{Network scheduling and dispatching:} \cite{Fendri2019} &
  \textbf{Scenario} - artificial: \cite{Peng2019}, \cite{Machado2018}, \cite{JAMRO201577}, \cite{Lei13}; realistic: \cite{Zhang2018a}, \cite{Xiang2014}, \cite{Fendri2019}.
\textbf{Data} - artificial:  \cite{Peng2019}, \cite{JAMRO201577}, \cite{Lei13}; realistic:  \cite{Machado2018}; real: \cite{Zhang2018a}, \cite{Xiang2014}, \cite{Fendri2019}. 
\textbf{Evaluation} - demonstration-simple:  \cite{Peng2019}, \cite{Machado2018}, \cite{JAMRO201577}, \cite{Lei13}; demonstration complex: \cite{Zhang2018a}, \cite{Xiang2014}, \cite{Fendri2019} \\ \hline
Security &
  \cite{RafaeldaSilva2017}, \cite{Tare2016}, \cite{zaitsev2016}, \cite{Chen2011}, \cite{li2018reliability}, \cite{xu2018petri}, \cite{Li2019a} &
  \textbf{Coloured:} \cite{RafaeldaSilva2017}, \cite{zaitsev2016}, \cite{Tare2016} \textbf{Stochastic:} \cite{li2018reliability}, \cite{Li2019a} \textbf{Timed:} \cite{xu2018petri}   \textbf{Hierarchical}: \cite{Chen2011} &
  \textbf{Network elements:} \cite{xu2018petri}, \textbf{Communication network:} \cite{zaitsev2016}, \textbf{Physical topology of SG:}  \cite{Tare2016}, \textbf{Communication services} \cite{RafaeldaSilva2017}, \textbf{Insiders:} \cite{Li2019a}, \textbf{Topology attacks:} \cite{li2018reliability}, \textbf{Cyber-physical attacks:} \cite{Chen2011} &
  \textbf{Scenario} - artificial: \cite{Chen2011}, \cite{Li2019a}, \cite{zaitsev2016}; realistic: \cite{li2018reliability}, \cite{RafaeldaSilva2017}, \cite{Tare2016}, \cite{xu2018petri}. \textbf{Data} - artificial: \cite{Chen2011}, \cite{Li2019a}, \cite{zaitsev2016}; realistic: \cite{li2018reliability}, \cite{RafaeldaSilva2017}, \cite{Tare2016}, \cite{xu2018petri}. \textbf{Evaluation} - demonstration-simple: \cite{Chen2011}, \cite{RafaeldaSilva2017}, \cite{Tare2016}, \cite{xu2018petri}, \cite{zaitsev2016}; demonstration complex: \cite{li2018reliability}, \cite{Li2019a} \\ \hline
\end{tabular}
\end{table*}

\vspace{0.7cm}
%-----------------
\begin{enumerate}
\item[\textbf{{RQ}2}.] \textit{Which types of Petri Nets are applied in smart grid domain?} 
\end{enumerate}

\begin{figure}[htb]
  \begin{center}
    \includegraphics[width=0.95\linewidth]{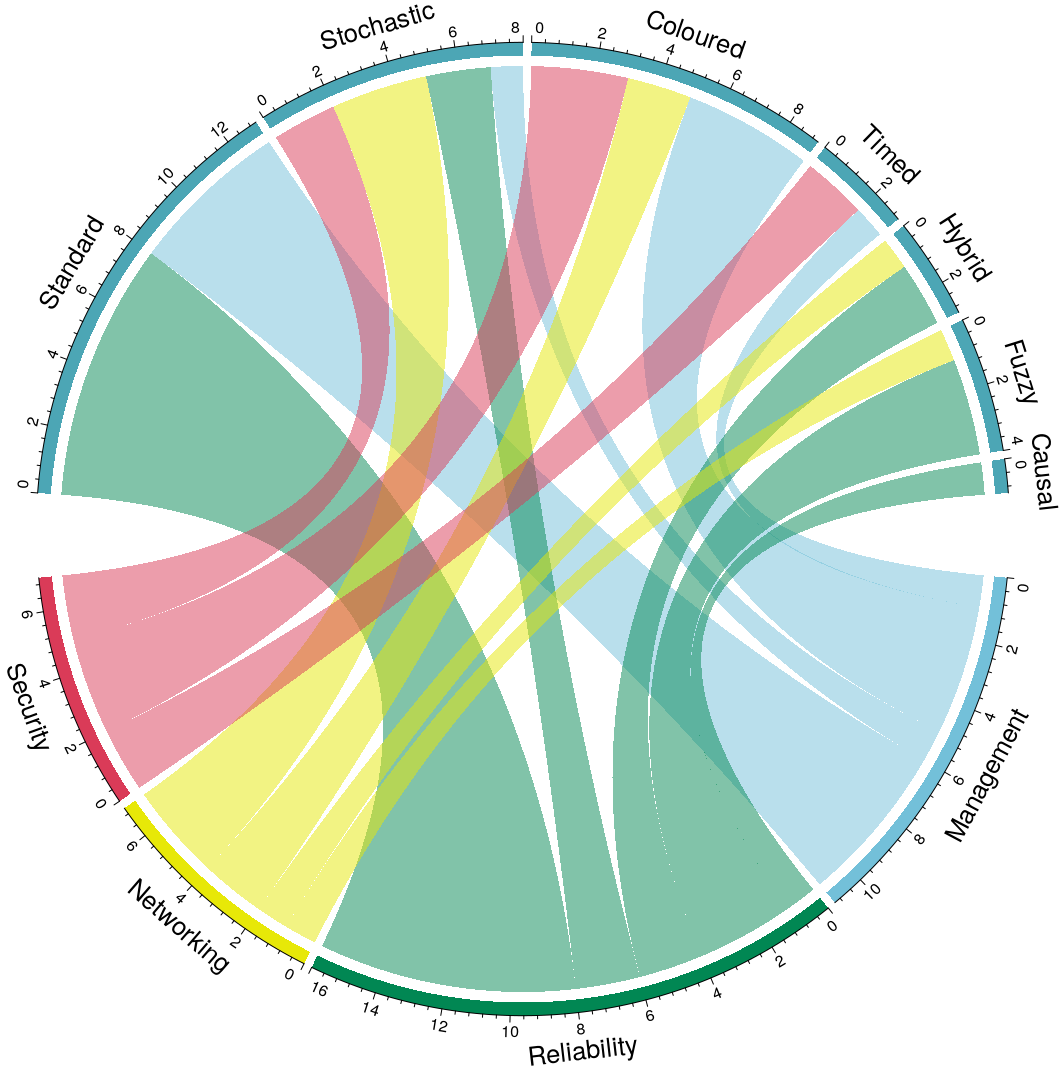}
  \end{center}
  \caption{Types of PNs applied by SG area}
  \label{fig:chord-diagram}
\end{figure}

We have classified the studies based on the type of PN adopted. We discuss them by SG's area and provide a representation in~Fig.\ref{fig:chord-diagram}.

\subsubsection*{Management} Standard PNs are used in this area covering aspects such as load control / bus voltage control in microgrids, energy management of power stations, but also the optimization of IT infrastructure for SGs \cite{Chamorro2012,Chamorro2012a,Kim2019,Kyriakarakos2012a,Lu2010,Zamani2011}. Colored PNs are used for modelling approaches to the reduction of power fluctuations in case of EVs, and modelling energy demand of different components \cite{Kaur2016,Zeineb2016}. Timed PNs are adopted for optimal energy scheduling algorithm for loads operation and for adapative autonomy of SGs to integrate experts opinions and environmental conditions \cite{Zamani2010}. Stochastic PNs are utilized for modelling optimization of acquired/consumed energy in smart homes \cite{Gentile2014}.

\subsubsection*{Networking} 
Stochastic PNs, Coloured PNs, Fuzzy PNs, Timed Hybrid PNs have been widely used to model and analyze complex networking systems. They allow us to have more insights for the performance and quality of service in network design and operation. Choosing a proper PN variant to address the specific issues in networking contexts. In networking, SPNs can be used to model systems with uncertain behaviors, such as network traffic, queuing systems, and fault tolerance \cite{Lei13,Peng2019,Xiang2014}. Also, CPNs are mostly used to represent the systems with heterogeneous components and complex protocols in networking \cite{Machado2018,JAMRO201577}. FPNs are usually applied to incorporate fuzzy logic to represent imprecise or uncertain information in the network modeling process \cite{Zhang2018a}. Finally, TPNs are valuable for modeling real-time systems such as communication networks with timing requirements \cite{Fendri2019}. Different PNs variants can be used for network design and optimization, it is critical to choose PNs to deal with complex network systems and scalability in Large-scale networks. 

\subsubsection*{Reliability} Standard PNs, Stochastic PNs, Fluid Stochastic PNs, Hybrid PNs, and Causal PNs have been used as reliability analysis tools for modelling system behaviors. Leveraging different PNs in reliability analysis can contribute to the development of dependable and resilient systems across different application domains. Standard PNs can be used to represent the dynamic interactions of system components, failure modes, and repair processes \cite{Saki2011,Liu2010,Calderaro2011,wang2015a,Liu2010,Calderaro2011,Panchal2016}. SPNs are usually used in modeling reliability scenarios with random events such as component failures and repair times \cite{Zeng2011,Zeng2012,Wang2014a,Morris2017,Mahdi2017}. Further, Fluid SPNs combine discrete events (e.g., component failures) and continuous behaviors (e.g., degradation processes) that impact reliability \cite{Huels2016,Marrone2016}. In line with Fluid SPNs, HPNs can model complex systems where digital controls interact with physical processes to assess system reliability \cite{Ghasemieh2015,Matos2016}. FPNs can be used to handle failure probabilities and assess system reliability under uncertainty\cite{Wang2014,Chen2015,Xu2019,Xu2019,Kiaei2020}. Causal PNs focus on the cause-and-effect relationships between events in the reliability analysis \cite{Diekhake2013}. They can be used to model fault propagation, dependencies among components, and the impact of failures from system behaviors. Overall, different PNs variants can be used to identify root causes of system failures, evaluating system availability, and ensuring the fault tolerance.

\subsubsection*{Security} Within the security area we have found out that type of PN used, while varying from reference to reference, follows a pattern, as four (4) of them use Coloured PNs \cite{RafaeldaSilva2017,Tare2016,zaitsev2016,Chen2011}, which means up to a 57\% of the papers related to security. It is interesting to mention that \cite{Chen2011} is the only reference of the area that mentions multiple types of PNs, which include, aside from the aforementioned Coloured PNs, Timed PNs, and Standard PNs. In this sense, \cite{xu2018petri} also uses Timed PNs. Finally, \cite{li2018reliability,Li2019a} make use of Stochastic PNs, amounting to a 28\% of the total, yet they constitute the second highest percentage of the area of security.

\vspace{0.7cm}
%-----------------
\begin{enumerate}
\item[\textbf{{RQ}3}.] \textit{What is modeled by Petri Nets in the identified areas?}  
\end{enumerate}

\subsubsection*{Management} PNs were used to model different components (producer, consumer, smart meter, energy management unit, local energy generator, utility) of a SG~\cite{Zeineb2016}. Similarly, Several PNs submodels were adopted (smart meter, energy model, accounting, smart appliances, solar panels, batteries)~\cite{Gentile2014} -- with transitions dealing with three levels: energy flux, energy data, accounting information. IF-THEN rules were proposed for the decision of Level of Automation (LOA)~\cite{Zamani2010}. The model was divided into primary rules and veto rules. PN nodes represent performance shaping factors (e.g., night/day, network age, load, network faults). In another work, employ PN models we employed for designing the decision model of the best composition of the SG infrastructure~\cite{Zamani2011}. One model determines the cost of the infrastructure, another the the performance. Control networks were also modelled with PNs~\cite{Chamorro2012a}. Transitions represent inputs and outputs could be continuous time, discrete time or event driven quantities, while nodes represent units that can be turned on/off. Similarly a PN model that comprises eight PN subnets interconnected through fusion places was adopted in~\cite{Munoz2010}. There can be a probability of failure e.g., when open/closing a circuit breaker. A PN model is used to set the preferred temperature in the building~\cite{Bouazza2019}. Nodes represent several occupancy states, transitions various conditions,like users detected in a room or not. PNs are used to model the event loop feeder element status information in~\cite{Congqi2013}. Nodes and transitions represent the hierarchy of states and events of an energy hub modelling a microgrid structure. In Kaur et al.~\cite{Kaur2016}, the PN model is used for balancing energy loads between charging stations and electric vehicles. A PN model represents the  states of a bus voltage control strategy in power flow computations in~\cite{Kim2019}. To model energy management of a photovoltaic-based power station, a PN model was implemented in~\cite{Lu2010}: nodes can represent a fully charged battery and transitions can represent the value of the state of charge.

\subsubsection*{Networking}
A variety of artifacts have been modelled in the network domain that covers the aspects of communication, resource optimization, and scheduling. First, network communication is to tackle the communication models for SG.\cite{JAMRO201577} uses systems modelling language diagrams to specify communication tasks and parameters to generate a Timed CPN model of the communication subsystem, and assess the performance and predict the behavior of the target system. Similarly, \cite{Lei13} solves how to analyze the Device-to-Device communication performance in smart grid. \cite{Xiang2014} proposed several backup strategies with different architectures for critical components for control center networks. Second, resource optimization in a network can be addressed to the performance analysis. \cite{Machado2018} investigates how to assist distribution system operators in the process of design and performance analysis of IEC 61850 networks in the microgrid context. \cite{Zhang2018a} proposes a model to operators in monitoring and controlling the real-time manufacturing process by the sensing of manufacturing information and the management of sensor. Third, network scheduling is focused on how to optimize the supply loads in SGs. \cite{Peng2019} contributes to the comprehensive production scheduling and energy profile assessment for sustainable re-manufacturing. Also, \citet{Fendri2019} proposes an energy flow scheduling algorithm for a PV/B chalet installation to assure continuous supply of loads.

\subsubsection*{Reliability}
In the reliability domain, there are two main artifacts that are modelled by PNs. The first artifact is failure diagnostics, monitoring and recovery. The failure diagnostics have been addressed from various aspects such by real-time data analysis~\cite{Saki2011}, fusion of multi-source data~\cite{Liu2010,Wang2014,Kiaei2020}, parity checking operations~\cite{Calderaro2011}, divided fault zones~\cite{Sreerama2017}, fault probability~\cite{Xu2019}, customers' consumption profiles~\cite{Chen2015}, decision sequences~\cite{Matos2016}. Failure causes can be identified by the monitoring. \cite{Diekhake2013} proposed an online monitoring of a distributed automation system for the recognition of message sequences. For failure recovery, \cite{Jiang2018} proposed a methodology to avoid the output interruption in case of a fault detection. It can be seen that the majority of the works are focused on failure diagnostics, whereas less works have been conducted on failure monitoring and recovery with PNs. The other artifact is reliability analysis, which also includes dependability analysis, risk analysis, availability analysis and resiliency analysis. Those analyses can be used to serve for the failure diagnostics, monitoring and recovery, where the research is focused on failure dynamics \cite{Panchal2016}, failure modes \cite{Mahdi2017}, and fault recovery techniques \cite{Morris2017}. Further, \citet{Zeng2011,Zeng2012} proposed a model for dependability analysis for control center networks. \cite{Huels2016} conducted the analysis between energy provider and the user and the system survivability. 

\subsubsection*{Security}
\cite{xu2018petri} models network elements, such as Firewall or Web serve, as well as the actions performed on them (Send email, Inject virus). On \cite{zaitsev2016} the elements modeled by the PN are, either a single computer or a network on which application software is installed, and how they produce produces and consumes packets that model the grid workload. Additionally, it implements the modelling of the attacks. The physical topology of the grid, including elements such as Master Station, Network, and Outstation, are modelled in \cite{Tare2016}, but it also includes the modelling of a cyber attack. The work done on \cite{RafaeldaSilva2017} focuses on using the international standard IEC 61850, which defines the communication and services form between different equipment present in the automation of power electrical systems, as the element to be modelled with PNs, but it is also extended with modelling an attack. \cite{Li2019a} focuses on modelling the insiders and their actions within the SG infrastructure but without going on detail about the gird itself. On \cite{li2018reliability} the elements modelled are topology attacks, as well as the cyber and physical parts of the SG. Finally, \cite{Chen2011} uses PNs to model cyber-physical attacks.

\vspace{0.7cm}

%-----------------
\begin{enumerate}
\item[\textbf{{RQ}4}.] \textit{What type of evaluation has been carried out for the use of the Petri Nets?} 
\end{enumerate}

We have further classified the evaluation into scenario, data, and validation levels. In the scenario, we have three levels that are \textit{artificial}, \textit{realistic}, and \textit{real-world} scenario, which correspond to the feasibility of the scenario across from \textit{"it is not designed with a real case"}, to \textit{"it can be simulated or adapted to a real case"}, and \textit{"it is based on a real case"}. This classification level is further used in the data component for the evaluation. For the validation, three levels are also specified: \textit{simple demonstration}, \textit{complex demonstration}, and \textit{comparison}. The \textit{simple demonstration} includes the validation with simple showcases as an evaluation, whereas the \textit{complex demonstration} contains more complex simulations such as sensitivity analysis. Finally, the \textit{validation with comparison} means the study has compared the proposed model with previous models and approaches. 

\subsubsection*{Management} There are several articles that build on top of artificial scenarios to evaluate the PN models proposed in the research area \cite{Gentile2014,Zamani2010,Zamani2011,Zeineb2016}. Other articles focus on more realistic scenarios \cite{Chamorro2012a,Munoz2010,Congqi2013,Kaur2016,Kim2019,Bouazza2019} -- while only one study covers a real scenario adopting the power production plan provided by the grid operator~\cite{Lu2010}. Many of the articles cite the solutions adopted for implementing the PN models and/or the simulations (e.g., Matlab and Platform Independent Petri Net (PIPE)~\cite{Bouazza2019}, DIgSILENT and MATPOWER~\cite{Kim2019}). However, our review found that no article provides the source code, while some provide the datasets adopted (e.g., ERCOT dataset~\cite{Kaur2016}). In most articles, there is either a simple or complex evaluation, while only ~\cite{Kaur2016} compares the model's results with the one proposed in ~\cite{liu2014vehicle}.

\subsubsection*{Networking} In the networking area, the evaluation scenarios are mostly artificial and realistic as proper simulations. For example, \cite{Peng2019,Machado2018,JAMRO201577,Lei13} have designed their own scenarios to validate the PN models or verify the accuracy of the analytical results. Other works in the networking such as \cite{Zhang2018a,Xiang2014,Fendri2019} are more focused on realistic simulations that can be further applied in practice. The data used in the evaluations are mostly artificial such as in \cite{Peng2019,JAMRO201577,Lei13} and real such as in  \cite{Zhang2018a,Xiang2014,Fendri2019}. This reflects that in the networking domain, we are able to obtain real-world data to conduct case studies or simulations. 

\subsubsection*{Reliability} In the reliability area, a large number of works such as \cite{Liu2010,Zeng2012,Wang2014,Panchal2016,Matos2016,Mahdi2017,Sreerama2017,Jiang2018} have just used artificial scenario and data. Thus, this also results in a simple validation. Some works such as \cite{Wang2014a,Marrone2016} are without evaluations or case studies. Some works with realistic scenario and data are validated based on open standard or systems. The evaluation in \cite{wang2015a} is based on IEEE 118-bus standard. \cite{Chen2015} used the IEEE 30-bus system. \cite{Morris2017} implemented Roy Billinton Test System. \cite{Xu2019} applied IEEE 14-bus power system in their evaluation.
Further, \cite{Kiaei2020} used IEEE 33 node feeder test system in their evaluation. Those works are more realistic and close to real implementation. There exist some works that used real scenarios or datasets in their evaluation. Those data can come from public datasets such as in \cite{Huels2016,Ghasemieh2015} or from real-world project and system \cite{Diekhake2013,Xu2019}. Furthermore, the evaluation in \cite{Ghasemieh2015} has compared different battery management and discharging strategies in the scenario of smart house survivability. The results indicate that the evaluations in reliability area need to consider more realistic and real scenario and data. Limited evaluation has considered to compared their work in the evaluation. 

\subsubsection*{Security} Based on the references considered for the security area, we have found that all those that had an artificial scenario, also have artificial data, as in \cite{Chen2011,Li2019a,zaitsev2016}. On the other hand, the other papers use realistic scenarios with realistic data, as in \cite{li2018reliability,RafaeldaSilva2017,Tare2016,xu2018petri}. This means that there are no real-world data or scenario. Regarding the evaluation carried out, those fall mostly under the category of a simple demonstration, as in the cases of \cite{Chen2011,RafaeldaSilva2017,Tare2016,xu2018petri,zaitsev2016}, with just \cite{li2018reliability,Li2019a} having a complex demonstration. There are no validations compare the results with other similar approaches. In general, it is easy to conclude that within the security area, the experiments and simulations made are of poor quality and have little to no replicability mostly because of the authors do not share the information that is required. Let it be noted, nonetheless, that with a high level of effort, some of the experiments could be replicated thanks to some extensive result data, but that is far from what is to be expected from a scientific experiment. 

\begin{figure}[htb]
  \begin{center}
     \includegraphics[width=0.8\linewidth]{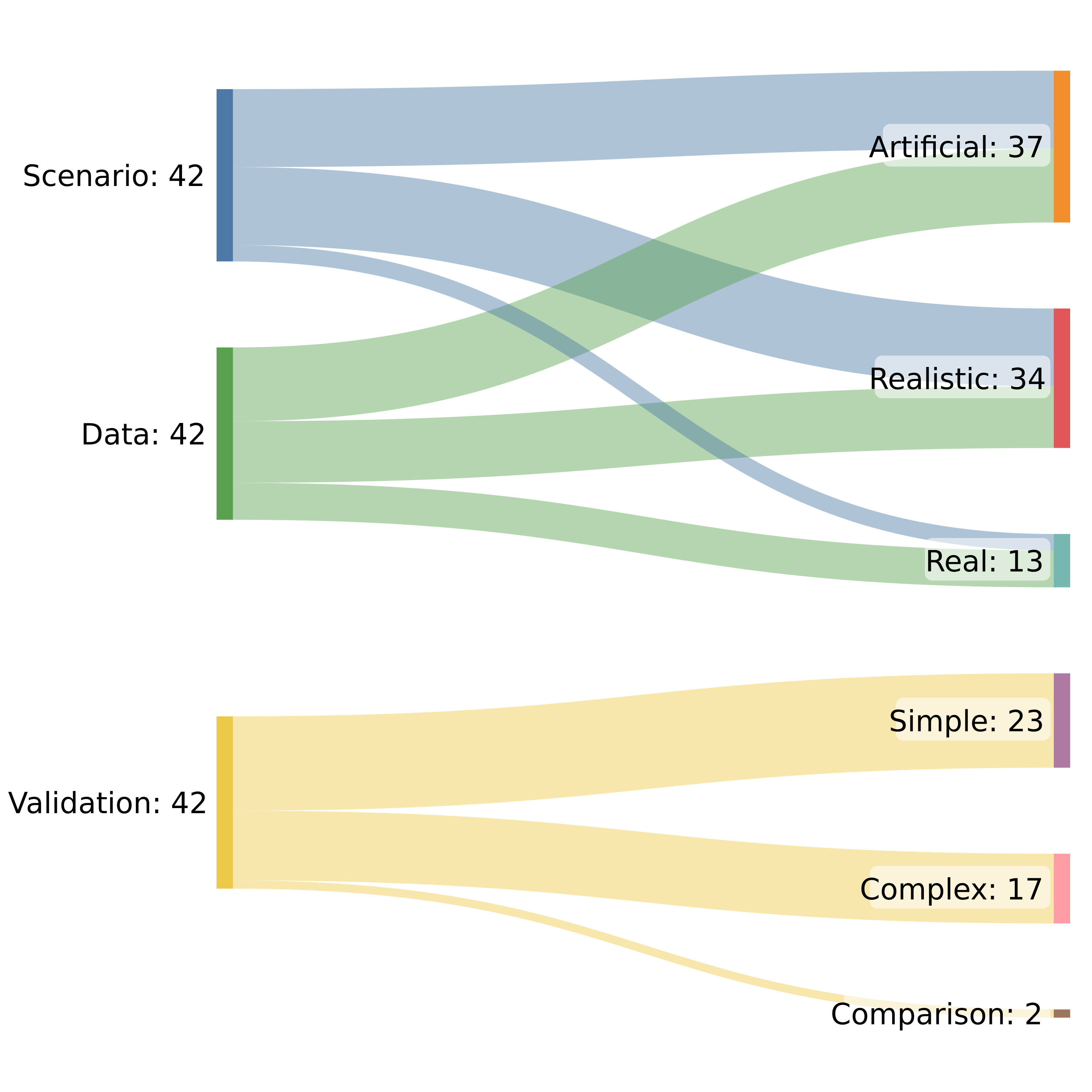}
  \end{center}
  \caption{Evaluation of Petri Nets papers across different areas in SGs 
  }
  \label{fig:across-diagram}
\end{figure}

To further interpret the results, we summarize the evaluations of PNs papers across the four identified areas in SGs as shown in Fig.~\ref{fig:across-diagram}. For the evaluation scenario, most PN models, regardless of the area, are validated within artificial and realistic scenarios. Instead of designing a scenario by the authors, many use the existing systems, such as IEEE bus systems, in the evaluations. It would be desirable that more real-world scenarios are to be included in the evaluation. 
Most artificial scenarios will result in artificial data. However, an artificial or a realistic scenario might use real-world datasets, although this is rare. With the recent development of open data, the use of public datasets in the evaluation is emerging. Also, it will become possible to compare different PN models further when the same datasets are used. Thus, this also shows limited comparison in the evaluations, especially the comparison among previous works rather than the comparison of different settings in a scenario. Based on our review, we found that if it is difficult to apply a real-world scenario, realistic simulations can be considered a good practice in the evaluation. Moreover, future evaluations may first consider public datasets in SGs and compare the results with benchmarks or other works.

\subsection{Threats to Validity}
There are several threats to validity that may potentially affect the reliability of this paper. Internal threats to validity are constituted mainly by selection bias and data extraction bias. It is expected that a scoping review cannot cover the whole research published in the area but rather define and go over a subset~\cite{munn2018systematic}. Furthermore, the topic of PNs in SGs crosses various disciplines, such as computer science, engineering, infrastructure, and energy management. Given the broad scope of the paper, the search process may oversee papers in specific domains. However, we have covered the widely used literature databases, especially from the computer science perspective. The data extraction is based on the authors' understanding and interpretation. To eliminate this bias, the authors have read the paper individually and confirmed the summary from each other. Regarding the external threats to validity, as the PNs in SGs are still an evolving topic, the insights derived from the review may not be generalizable from one domain to another. Thus, we conclude that the findings are inside an identified area. Also, to mitigate the temporal heterogeneity, the paper selection is focused on papers from the last two decades. Due to a specific search scope across different literature databases, we do not include papers from the last two years. Given this limitation, we believe this will not influence the reliability of the derived results because new papers mostly use the same methodologies from previous papers.

\section{Conclusion} 
\label{sec:conclusion}

The work done on this paper shows the relation between PNs and SGs. Specifically, we have looked in depth at more than 40 references that go over the usage of PNs in the SGs domain. While looking at these papers, we have also been able to locate four different areas of vital importance for SGs on which PNs have applied: Management, Reliability, Networking, and Security. Within these areas, we had the opportunity to see not only what the PNs were being used for, but also what aspects were modelled, and the evaluation that was carried out. All this has been the main effort to answer the research questions we proposed in Section~\ref{sec:research-methods}.

Based on the review, we can extract some conclusions. The first is that the use of PNs in the domain of SGs has been focused more on the reliability area and management area, while the effort done on the networking and security areas is of less extent. Following this trend, the evaluations that have been performed on the first two areas are more in-depth than the ones of the other two, including more realistic scenarios and, definitely, more complex demonstrations. This finding can be supported by the only two evaluations by comparison that appear in those areas. Nevertheless, it is also interesting that the type of PNs more used in those areas is the one of Standard PN, while in the other areas, the type of PN is more distributed among more complex types. This aspect could indicate that the extra literature found in the management and reliability areas could be due to a much easier extraction of results because of the adoption of more standard methods. One reason can also be due to the maturity of the research areas as more recent efforts are placed into security and network communication.

Overall, we have provided a comprehensive overview of the current status of intertwining between PNs and SGs. This paper is to serve as the basis for the understanding of the current application of PNs to the SGs domain to see which areas are covered and which research directions have been covered over the years. As expected based on previous research, the research points directly toward the usefulness of PNs in the domain of SGs, which have been applied to model many aspects, from power consumption to cyber-physical attacks, proving their flexibility and broad applicability.

\begin{acks}
The work was supported by ERDF "CyberSecurity, CyberCrime and Critical Information Infrastructures Center of Excellence" (No. CZ.02.1.01/0.0/0.0/16\_019/0000822).
\\
This work was partially supported by the SUNRISE project, funded from the Horizon Europe research programme (2021-2027) under grant agreement no. 101073821.
\end{acks}

\balance

\bibliographystyle{ACM-Reference-Format}
\input{PN24.bbl}

%\bibliography{ref} 

\end{document}

%% file: PN24.bbl

%% file: PN24.bbl
\begin{thebibliography}{90}

%%% the Web address.

\ifx \showCODEN    \undefined \def \showCODEN     #1{\unskip}     \fi
\ifx \showDOI      \undefined \def \showDOI       #1{#1}\fi
\ifx \showISBNx    \undefined \def \showISBNx     #1{\unskip}     \fi
\ifx \showISBNxiii \undefined \def \showISBNxiii  #1{\unskip}     \fi
\ifx \showISSN     \undefined \def \showISSN      #1{\unskip}     \fi
\ifx \showLCCN     \undefined \def \showLCCN      #1{\unskip}     \fi
\ifx \shownote     \undefined \def \shownote      #1{#1}          \fi
\ifx \showarticletitle \undefined \def \showarticletitle #1{#1}   \fi
\ifx \showURL      \undefined \def \showURL       {\relax}        \fi
% invisible to TeX
\providecommand\bibfield[2]{#2}
\providecommand\bibinfo[2]{#2}
\providecommand\natexlab[1]{#1}
\providecommand\showeprint[2][]{arXiv:#2}

\bibitem[\protect\citeauthoryear{Ahmad, Ghadi, Adnan, and Ali}{Ahmad
  et~al\mbox{.}}{2023a}]%
        {AhmadGAA23}
\bibfield{author}{\bibinfo{person}{Naqash Ahmad}, \bibinfo{person}{Yazeed~Yasin
  Ghadi}, \bibinfo{person}{Muhammad Adnan}, {and} \bibinfo{person}{Mansoor
  Ali}.} \bibinfo{year}{2023}\natexlab{a}.
\newblock \showarticletitle{From Smart Grids to Super Smart Grids: {A} Roadmap
  for Strategic Demand Management for Next Generation {SAARC} and European
  Power Infrastructure}.
\newblock \bibinfo{journal}{\emph{{IEEE} Access}}  \bibinfo{volume}{11}
  (\bibinfo{year}{2023}), \bibinfo{pages}{12303--12341}.
\newblock


\bibitem[\protect\citeauthoryear{Ahmad, Shafiullah, Ahmed, and
  Alowaifeer}{Ahmad et~al\mbox{.}}{2023b}]%
        {AhmadSAA23}
\bibfield{author}{\bibinfo{person}{Saad Ahmad}, \bibinfo{person}{Md
  Shafiullah}, \bibinfo{person}{Chokri~Belhaj Ahmed}, {and}
  \bibinfo{person}{Maad Alowaifeer}.} \bibinfo{year}{2023}\natexlab{b}.
\newblock \showarticletitle{A Review of Microgrid Energy Management and Control
  Strategies}.
\newblock \bibinfo{journal}{\emph{{IEEE} Access}}  \bibinfo{volume}{11}
  (\bibinfo{year}{2023}), \bibinfo{pages}{21729--21757}.
\newblock


\bibitem[\protect\citeauthoryear{Alahakoon and Yu}{Alahakoon and Yu}{2016}]%
        {ref:alahakoon2016futuresurvey}
\bibfield{author}{\bibinfo{person}{Damminda Alahakoon} {and}
  \bibinfo{person}{Xinghuo Yu}.} \bibinfo{year}{2016}\natexlab{}.
\newblock \showarticletitle{Smart electricity meter data intelligence for
  future energy systems: A survey}.
\newblock \bibinfo{journal}{\emph{IEEE Transactions on Industrial Informatics}}
  \bibinfo{volume}{12}, \bibinfo{number}{1} (\bibinfo{year}{2016}),
  \bibinfo{pages}{425--436}.
\newblock


\bibitem[\protect\citeauthoryear{Benetti, Caprino, Della~Vedova, and
  Facchinetti}{Benetti et~al\mbox{.}}{2016}]%
        {ref:benetti2016slr}
\bibfield{author}{\bibinfo{person}{Guido Benetti}, \bibinfo{person}{Davide
  Caprino}, \bibinfo{person}{Marco~L Della~Vedova}, {and}
  \bibinfo{person}{Tullio Facchinetti}.} \bibinfo{year}{2016}\natexlab{}.
\newblock \showarticletitle{Electric load management approaches for peak load
  reduction: A systematic literature review and state of the art}.
\newblock \bibinfo{journal}{\emph{Sustainable Cities and Society}}
  \bibinfo{volume}{20} (\bibinfo{year}{2016}), \bibinfo{pages}{124--141}.
\newblock


\bibitem[\protect\citeauthoryear{Bhardwaj and Singh}{Bhardwaj and
  Singh}{2017}]%
        {ref:bhardwaj2017smartgridsanalyticssystematic}
\bibfield{author}{\bibinfo{person}{Ashu Bhardwaj} {and}
  \bibinfo{person}{Williamjeet Singh}.} \bibinfo{year}{2017}\natexlab{}.
\newblock \showarticletitle{Systematic Review of Smart Grid Analytics}.
\newblock \bibinfo{journal}{\emph{International Journal of Advanced Research in
  Computer Science}} \bibinfo{volume}{8}, \bibinfo{number}{5}
  (\bibinfo{year}{2017}), \bibinfo{pages}{2333--2338}.
\newblock


\bibitem[\protect\citeauthoryear{Bouazza and Deabes}{Bouazza and
  Deabes}{2019}]%
        {Bouazza2019}
\bibfield{author}{\bibinfo{person}{Kheir~Eddine Bouazza} {and}
  \bibinfo{person}{Wael~A Deabes}.} \bibinfo{year}{2019}\natexlab{}.
\newblock \showarticletitle{{Smart Petri Nets Temperature Control Framework for
  Reducing Building Energy Consumption}}.
\newblock \bibinfo{journal}{\emph{Sensors}} \bibinfo{volume}{19},
  \bibinfo{number}{11} (\bibinfo{year}{2019}), \bibinfo{pages}{2441}.
\newblock
\urldef\tempurl%
\url{https://doi.org/10.3390/s19112441}
\showDOI{\tempurl}


\bibitem[\protect\citeauthoryear{Buccafurri, Musarella, and Nardone}{Buccafurri
  et~al\mbox{.}}{2019}]%
        {Buccafurri2019}
\bibfield{author}{\bibinfo{person}{Francesco Buccafurri},
  \bibinfo{person}{Lorenzo Musarella}, {and} \bibinfo{person}{Roberto
  Nardone}.} \bibinfo{year}{2019}\natexlab{}.
\newblock \showarticletitle{{A routing algorithm increasing the transmission
  availability in smart grids}}.
\newblock \bibinfo{journal}{\emph{Simulation Series}}
  \bibinfo{volume}{2019-July}, Article \bibinfo{articleno}{47}
  (\bibinfo{year}{2019}), \bibinfo{numpages}{12}~pages.
\newblock
\showISSN{07359276}


\bibitem[\protect\citeauthoryear{Cabasino, Giua, and Seatzu}{Cabasino
  et~al\mbox{.}}{2014}]%
        {HPN}
\bibfield{author}{\bibinfo{person}{Maria Cabasino}, \bibinfo{person}{Alessandro
  Giua}, {and} \bibinfo{person}{Carla Seatzu}.}
  \bibinfo{year}{2014}\natexlab{}.
\newblock \bibinfo{booktitle}{\emph{Modelling Manufacturing Systems and
  Inventory Control Systems with Hybrid Petri Nets}}.
\newblock \bibinfo{publisher}{CRC Press}, \bibinfo{pages}{75--103}.
\newblock
\showISBNx{978-1-4665-6155-7}


\bibitem[\protect\citeauthoryear{Calderaro, Hadjicostis, Piccolo, and
  Siano}{Calderaro et~al\mbox{.}}{2011}]%
        {Calderaro2011}
\bibfield{author}{\bibinfo{person}{Vito Calderaro},
  \bibinfo{person}{Christoforos~N Hadjicostis}, \bibinfo{person}{Antonio
  Piccolo}, {and} \bibinfo{person}{Pierluigi Siano}.}
  \bibinfo{year}{2011}\natexlab{}.
\newblock \showarticletitle{{Failure Identification in Smart Grids Based on
  Petri Net Modeling}}.
\newblock \bibinfo{journal}{\emph{IEEE Transactions on Industrial Electronics}}
  \bibinfo{volume}{58}, \bibinfo{number}{10} (\bibinfo{year}{2011}),
  \bibinfo{pages}{4613--4623}.
\newblock
\showISSN{0278-0046}
\urldef\tempurl%
\url{https://doi.org/10.1109/TIE.2011.2109335}
\showDOI{\tempurl}


\bibitem[\protect\citeauthoryear{Castellanos~Contreras and
  Rodríguez~Urrego}{Castellanos~Contreras and Rodríguez~Urrego}{2023}]%
        {ref:Contreras2023}
\bibfield{author}{\bibinfo{person}{Jose~Ulises Castellanos~Contreras} {and}
  \bibinfo{person}{Leonardo Rodríguez~Urrego}.}
  \bibinfo{year}{2023}\natexlab{}.
\newblock \showarticletitle{Technological Developments in Control Models Using
  Petri Nets for Smart Grids: A Review}.
\newblock \bibinfo{journal}{\emph{Energies}} \bibinfo{volume}{16},
  \bibinfo{number}{8} (\bibinfo{year}{2023}), 21.
\newblock
\urldef\tempurl%
\url{https://doi.org/10.3390/en16083541}
\showDOI{\tempurl}


\bibitem[\protect\citeauthoryear{Chamorro and Jimenez}{Chamorro and
  Jimenez}{2012}]%
        {Chamorro2012a}
\bibfield{author}{\bibinfo{person}{H~R Chamorro} {and} \bibinfo{person}{J~F
  Jimenez}.} \bibinfo{year}{2012}\natexlab{}.
\newblock \showarticletitle{{Use of petri nets for load sharing control in
  distributed generation applications}}. In \bibinfo{booktitle}{\emph{2012 3rd
  IEEE International Symposium on Power Electronics for Distributed Generation
  Systems (PEDG)}}. \bibinfo{publisher}{IEEE}, \bibinfo{pages}{731--736}.
\newblock
\showISSN{2329-5767}
\urldef\tempurl%
\url{https://doi.org/10.1109/PEDG.2012.6254083}
\showDOI{\tempurl}


\bibitem[\protect\citeauthoryear{Chamorro, Ordonez, and Jimenez}{Chamorro
  et~al\mbox{.}}{2012}]%
        {Chamorro2012}
\bibfield{author}{\bibinfo{person}{H~R Chamorro}, \bibinfo{person}{C~A
  Ordonez}, {and} \bibinfo{person}{J~F Jimenez}.}
  \bibinfo{year}{2012}\natexlab{}.
\newblock \showarticletitle{{Coordinated control based Petri Nets for
  Microgrids including wind farms}}. In \bibinfo{booktitle}{\emph{2012 IEEE
  Power Electronics and Machines in Wind Applications}}.
  \bibinfo{publisher}{IEEE}, \bibinfo{pages}{1--6}.
\newblock
\urldef\tempurl%
\url{https://doi.org/10.1109/PEMWA.2012.6316370}
\showDOI{\tempurl}


\bibitem[\protect\citeauthoryear{Chen, Zhan, Huang, Chen, and Lin}{Chen
  et~al\mbox{.}}{2015}]%
        {Chen2015}
\bibfield{author}{\bibinfo{person}{S Chen}, \bibinfo{person}{T Zhan},
  \bibinfo{person}{C Huang}, \bibinfo{person}{J Chen}, {and} \bibinfo{person}{C
  Lin}.} \bibinfo{year}{2015}\natexlab{}.
\newblock \showarticletitle{{Nontechnical Loss and Outage Detection Using
  Fractional-Order Self-Synchronization Error-Based Fuzzy Petri Nets in
  Micro-Distribution Systems}}.
\newblock \bibinfo{journal}{\emph{IEEE Transactions on Smart Grid}}
  \bibinfo{volume}{6}, \bibinfo{number}{1} (\bibinfo{year}{2015}),
  \bibinfo{pages}{411--420}.
\newblock
\showISSN{1949-3061}
\urldef\tempurl%
\url{https://doi.org/10.1109/TSG.2014.2345780}
\showDOI{\tempurl}


\bibitem[\protect\citeauthoryear{Chen, Sanchez-Aarnoutse, Buford,
  S{\'{a}}nchez-Aarnoutse, and Buford}{Chen et~al\mbox{.}}{2011}]%
        {Chen2011}
\bibfield{author}{\bibinfo{person}{Thomas~M Chen}, \bibinfo{person}{J~C
  Sanchez-Aarnoutse}, \bibinfo{person}{John~F Buford},
  \bibinfo{person}{Juan~Carlos S{\'{a}}nchez-Aarnoutse}, {and}
  \bibinfo{person}{John~F Buford}.} \bibinfo{year}{2011}\natexlab{}.
\newblock \showarticletitle{{Petri Net Modeling of Cyber-Physical Attacks on
  Smart Grid}}.
\newblock \bibinfo{journal}{\emph{IEEE Transactions on Smart Grid}}
  \bibinfo{volume}{2}, \bibinfo{number}{4} (\bibinfo{year}{2011}),
  \bibinfo{pages}{741--749}.
\newblock
\showISSN{1949-3061}
\urldef\tempurl%
\url{https://doi.org/10.1109/TSG.2011.2160000}
\showDOI{\tempurl}


\bibitem[\protect\citeauthoryear{Cintuglu, Mohammed, Akkaya, and
  Uluagac}{Cintuglu et~al\mbox{.}}{2017}]%
        {ref:cintuglu2017survey}
\bibfield{author}{\bibinfo{person}{Mehmet~Hazar Cintuglu},
  \bibinfo{person}{Osama~A Mohammed}, \bibinfo{person}{Kemal Akkaya}, {and}
  \bibinfo{person}{A~Selcuk Uluagac}.} \bibinfo{year}{2017}\natexlab{}.
\newblock \showarticletitle{A survey on smart grid cyber-physical system
  testbeds}.
\newblock \bibinfo{journal}{\emph{IEEE Communications Surveys \& Tutorials}}
  \bibinfo{volume}{19}, \bibinfo{number}{1} (\bibinfo{year}{2017}),
  \bibinfo{pages}{446--464}.
\newblock
\showISSN{1553-877X}
\urldef\tempurl%
\url{https://doi.org/10.1109/COMST.2016.2627399}
\showDOI{\tempurl}


\bibitem[\protect\citeauthoryear{Congqi, Tianmei, Runnan, and Guizhi}{Congqi
  et~al\mbox{.}}{2013}]%
        {Congqi2013}
\bibfield{author}{\bibinfo{person}{X Congqi}, \bibinfo{person}{L Tianmei},
  \bibinfo{person}{D Runnan}, {and} \bibinfo{person}{J Guizhi}.}
  \bibinfo{year}{2013}\natexlab{}.
\newblock \showarticletitle{{Research on Energy-Hub Control Method of
  Micro-grid Based on Multi-agent Petri Nets}}. In
  \bibinfo{booktitle}{\emph{2013 Fourth International Conference on Intelligent
  Systems Design and Engineering Applications}}. \bibinfo{pages}{468--471}.
\newblock
\urldef\tempurl%
\url{https://doi.org/10.1109/ISDEA.2013.512}
\showDOI{\tempurl}


\bibitem[\protect\citeauthoryear{{De Benedictis}, Mazzocca, Nardone, and
  Venticinque}{{De Benedictis} et~al\mbox{.}}{2018}]%
        {DeBenedictis2018}
\bibfield{author}{\bibinfo{person}{A {De Benedictis}}, \bibinfo{person}{N
  Mazzocca}, \bibinfo{person}{R Nardone}, {and} \bibinfo{person}{S
  Venticinque}.} \bibinfo{year}{2018}\natexlab{}.
\newblock \showarticletitle{{A Model-Based Evaluation Methodology for Smart
  Energy Systems}}. In \bibinfo{booktitle}{\emph{2018 IEEE International
  Conference on Smart Computing (SMARTCOMP)}}. \bibinfo{pages}{187--194}.
\newblock
\urldef\tempurl%
\url{https://doi.org/10.1109/SMARTCOMP.2018.00090}
\showDOI{\tempurl}


\bibitem[\protect\citeauthoryear{Diamantoulakis, Kapinas, and
  Karagiannidis}{Diamantoulakis et~al\mbox{.}}{2015}]%
        {ref:diamantoulakis2015big}
\bibfield{author}{\bibinfo{person}{Panagiotis~D Diamantoulakis},
  \bibinfo{person}{Vasileios~M Kapinas}, {and} \bibinfo{person}{George~K
  Karagiannidis}.} \bibinfo{year}{2015}\natexlab{}.
\newblock \showarticletitle{Big data analytics for dynamic energy management in
  smart grids}.
\newblock \bibinfo{journal}{\emph{Big Data Research}} \bibinfo{volume}{2},
  \bibinfo{number}{3} (\bibinfo{year}{2015}), \bibinfo{pages}{94--101}.
\newblock


\bibitem[\protect\citeauthoryear{Diekhake and Schnieder}{Diekhake and
  Schnieder}{2013}]%
        {Diekhake2013}
\bibfield{author}{\bibinfo{person}{Patrick Diekhake} {and}
  \bibinfo{person}{Eckehard Schnieder}.} \bibinfo{year}{2013}\natexlab{}.
\newblock \showarticletitle{{Online monitoring of a distributed building
  automation system to verify large sequences of bus messages by causal Petri
  net models}}.
\newblock \bibinfo{journal}{\emph{IECON Proceedings (Industrial Electronics
  Conference)}} (\bibinfo{year}{2013}), \bibinfo{pages}{3651--3655}.
\newblock
\showISBNx{9781479902248}
\urldef\tempurl%
\url{https://doi.org/10.1109/IECON.2013.6699716}
\showDOI{\tempurl}


\bibitem[\protect\citeauthoryear{Farhangi}{Farhangi}{2009}]%
        {farhangi2009path}
\bibfield{author}{\bibinfo{person}{Hassan Farhangi}.}
  \bibinfo{year}{2009}\natexlab{}.
\newblock \showarticletitle{The path of the smart grid}.
\newblock \bibinfo{journal}{\emph{IEEE power and energy magazine}}
  \bibinfo{volume}{8}, \bibinfo{number}{1} (\bibinfo{year}{2009}),
  \bibinfo{pages}{18--28}.
\newblock


\bibitem[\protect\citeauthoryear{Fendri and Chaabene}{Fendri and
  Chaabene}{2019}]%
        {Fendri2019}
\bibfield{author}{\bibinfo{person}{Dalia Fendri} {and} \bibinfo{person}{Maher
  Chaabene}.} \bibinfo{year}{2019}\natexlab{}.
\newblock \showarticletitle{{Hybrid Petri Net scheduling model of household
  appliances for optimal renewable energy dispatching}}.
\newblock \bibinfo{journal}{\emph{Sustainable Cities and Society}}
  \bibinfo{volume}{45} (\bibinfo{year}{2019}), \bibinfo{pages}{151--158}.
\newblock
\showISSN{2210-6707}
\urldef\tempurl%
\url{https://doi.org/10.1016/j.scs.2018.11.032}
\showDOI{\tempurl}


\bibitem[\protect\citeauthoryear{Fendri and Chaabene}{Fendri and
  Chaabene}{2020}]%
        {Fendri2020a}
\bibfield{author}{\bibinfo{person}{Dalia Fendri} {and} \bibinfo{person}{Maher
  Chaabene}.} \bibinfo{year}{2020}\natexlab{}.
\newblock \showarticletitle{{Application of Hybrid Petri Nets for the Energy
  Dispatching of an Isolated Micro-Grid}}.
\newblock \bibinfo{journal}{\emph{Int. J. of Applied Metaheuristic Computing}}
  \bibinfo{volume}{11}, \bibinfo{number}{1} (\bibinfo{year}{2020}),
  \bibinfo{pages}{61--72}.
\newblock
\urldef\tempurl%
\url{https://doi.org/10.4018/IJAMC.2020010105}
\showDOI{\tempurl}


\bibitem[\protect\citeauthoryear{Fereidunian, Zamani, {Sharifi K.}, and
  Lesani}{Fereidunian et~al\mbox{.}}{2011}]%
        {Fereidunian2011}
\bibfield{author}{\bibinfo{person}{A Fereidunian}, \bibinfo{person}{M~A
  Zamani}, \bibinfo{person}{M~A {Sharifi K.}}, {and} \bibinfo{person}{H
  Lesani}.} \bibinfo{year}{2011}\natexlab{}.
\newblock \showarticletitle{{AAHPNES: A Hierarchical Petri Net Expert System
  realization of adaptive autonomy in Smart Grid}}. In
  \bibinfo{booktitle}{\emph{2011 IEEE Trondheim PowerTech}}.
  \bibinfo{pages}{1--7}.
\newblock
\urldef\tempurl%
\url{https://doi.org/10.1109/PTC.2011.6019387}
\showDOI{\tempurl}


\bibitem[\protect\citeauthoryear{Ge, Bangui, and Buhnova}{Ge
  et~al\mbox{.}}{2018}]%
        {ref:ge2018big}
\bibfield{author}{\bibinfo{person}{Mouzhi Ge}, \bibinfo{person}{Hind Bangui},
  {and} \bibinfo{person}{Barbora Buhnova}.} \bibinfo{year}{2018}\natexlab{}.
\newblock \showarticletitle{Big Data for Internet of Things: A Survey}.
\newblock \bibinfo{journal}{\emph{Future Generation Computer Systems}}
  (\bibinfo{year}{2018}).
\newblock


\bibitem[\protect\citeauthoryear{Gentile, Marrone, Mazzocca, and
  Nardone}{Gentile et~al\mbox{.}}{2014}]%
        {Gentile2014}
\bibfield{author}{\bibinfo{person}{U Gentile}, \bibinfo{person}{S Marrone},
  \bibinfo{person}{N Mazzocca}, {and} \bibinfo{person}{R Nardone}.}
  \bibinfo{year}{2014}\natexlab{}.
\newblock \showarticletitle{{A Cost-Energy Trade-Off Model in Smart Energy
  Grids}}. In \bibinfo{booktitle}{\emph{2014 Ninth International Conference on
  P2P, Parallel, Grid, Cloud and Internet Computing}}.
  \bibinfo{pages}{394--399}.
\newblock
\urldef\tempurl%
\url{https://doi.org/10.1109/3PGCIC.2014.83}
\showDOI{\tempurl}


\bibitem[\protect\citeauthoryear{Ghasemieh, Haverkort, Jongerden, and
  Remke}{Ghasemieh et~al\mbox{.}}{2015}]%
        {Ghasemieh2015}
\bibfield{author}{\bibinfo{person}{Hamed Ghasemieh},
  \bibinfo{person}{Boudewijn~R. Haverkort}, \bibinfo{person}{Marijn~R.
  Jongerden}, {and} \bibinfo{person}{Anne Remke}.}
  \bibinfo{year}{2015}\natexlab{}.
\newblock \showarticletitle{{Energy Resilience Modelling for Smart Houses}}.
\newblock \bibinfo{journal}{\emph{Proceedings of the International Conference
  on Dependable Systems and Networks}}  \bibinfo{volume}{2015-Septe}
  (\bibinfo{year}{2015}), \bibinfo{pages}{275--286}.
\newblock
\showISBNx{9781479986293}
\urldef\tempurl%
\url{https://doi.org/10.1109/DSN.2015.31}
\showDOI{\tempurl}


\bibitem[\protect\citeauthoryear{Ghazi}{Ghazi}{2016}]%
        {mouzhireadinprogress}
\bibfield{author}{\bibinfo{person}{Doustmohammadi Ghazi, Z.}}
  \bibinfo{year}{2016}\natexlab{}.
\newblock \showarticletitle{Fault detection and power distribution optimization
  of smart grids based on hybrid Petri net}.
\newblock \bibinfo{journal}{\emph{Energy Systems}}  \bibinfo{volume}{8}
  (\bibinfo{year}{2016}), \bibinfo{pages}{465--493}.
\newblock


\bibitem[\protect\citeauthoryear{Gong, Zhang, Yang, Li, Liu, and Yao}{Gong
  et~al\mbox{.}}{2021}]%
        {GongZYLLY21}
\bibfield{author}{\bibinfo{person}{Ling{-}lin Gong}, \bibinfo{person}{Yizhuo
  Zhang}, \bibinfo{person}{Minghao Yang}, \bibinfo{person}{Yujia Li},
  \bibinfo{person}{Fang Liu}, {and} \bibinfo{person}{Qi Yao}.}
  \bibinfo{year}{2021}\natexlab{}.
\newblock \showarticletitle{A Review of Reliability, Vulnerability and
  Resilience Analysis of Smart Grid Based on Complex Network}. In
  \bibinfo{booktitle}{\emph{21st {IEEE} International Conference on Software
  Quality, Reliability and Security, {QRS} 2021 - Companion, Hainan, China,
  December 6-10, 2021}}. \bibinfo{publisher}{{IEEE}},
  \bibinfo{pages}{117--126}.
\newblock


\bibitem[\protect\citeauthoryear{Hu and Vasilakos}{Hu and Vasilakos}{2016}]%
        {ref:hu2016bigdatachallenges}
\bibfield{author}{\bibinfo{person}{Jiankun Hu} {and}
  \bibinfo{person}{Athanasios~V Vasilakos}.} \bibinfo{year}{2016}\natexlab{}.
\newblock \showarticletitle{Energy big data analytics and security: challenges
  and opportunities}.
\newblock \bibinfo{journal}{\emph{IEEE Transactions on Smart Grid}}
  \bibinfo{volume}{7}, \bibinfo{number}{5} (\bibinfo{year}{2016}),
  \bibinfo{pages}{2423--2436}.
\newblock


\bibitem[\protect\citeauthoryear{H{\"{u}}els and Remke}{H{\"{u}}els and
  Remke}{2016}]%
        {Huels2016}
\bibfield{author}{\bibinfo{person}{J H{\"{u}}els} {and} \bibinfo{person}{A
  Remke}.} \bibinfo{year}{2016}\natexlab{}.
\newblock \showarticletitle{{Energy Storage in Smart Homes: Grid-Convenience
  Versus Self-Use and Survivability}}. In \bibinfo{booktitle}{\emph{2016 IEEE
  24th International Symposium on Modeling, Analysis and Simulation of Computer
  and Telecommunication Systems (MASCOTS)}}. \bibinfo{pages}{385--390}.
\newblock
\showISSN{2375-0227}
\urldef\tempurl%
\url{https://doi.org/10.1109/MASCOTS.2016.33}
\showDOI{\tempurl}


\bibitem[\protect\citeauthoryear{Hujsa and Devillers}{Hujsa and
  Devillers}{2018}]%
        {HujsaD18}
\bibfield{author}{\bibinfo{person}{Thomas Hujsa} {and}
  \bibinfo{person}{Raymond~R. Devillers}.} \bibinfo{year}{2018}\natexlab{}.
\newblock \showarticletitle{On Deadlockability, Liveness and Reversibility in
  Subclasses of Weighted Petri Nets}.
\newblock \bibinfo{journal}{\emph{Fundam. Informaticae}} \bibinfo{volume}{161},
  \bibinfo{number}{4} (\bibinfo{year}{2018}), \bibinfo{pages}{383--421}.
\newblock


\bibitem[\protect\citeauthoryear{Jamro, Rzonca, and Rząsa}{Jamro
  et~al\mbox{.}}{2015}]%
        {JAMRO201577}
\bibfield{author}{\bibinfo{person}{Marcin Jamro}, \bibinfo{person}{Dariusz
  Rzonca}, {and} \bibinfo{person}{Wojciech Rząsa}.}
  \bibinfo{year}{2015}\natexlab{}.
\newblock \showarticletitle{Testing communication tasks in distributed control
  systems with SysML and Timed Colored Petri Nets model}.
\newblock \bibinfo{journal}{\emph{Computers in Industry}}  \bibinfo{volume}{71}
  (\bibinfo{year}{2015}), \bibinfo{pages}{77--87}.
\newblock
\showISSN{0166-3615}


\bibitem[\protect\citeauthoryear{Jeng and DiCesare}{Jeng and DiCesare}{1993}]%
        {JengD93}
\bibfield{author}{\bibinfo{person}{MuDer Jeng} {and} \bibinfo{person}{Frank
  DiCesare}.} \bibinfo{year}{1993}\natexlab{}.
\newblock \showarticletitle{A review of synthesis techniques for Petri nets
  with applications to automated manufacturing systems}.
\newblock \bibinfo{journal}{\emph{{IEEE} Trans. Syst. Man Cybern.}}
  \bibinfo{volume}{23}, \bibinfo{number}{1} (\bibinfo{year}{1993}),
  \bibinfo{pages}{301--312}.
\newblock


\bibitem[\protect\citeauthoryear{Jensen}{Jensen}{1996}]%
        {CPN}
\bibfield{author}{\bibinfo{person}{Kurt Jensen}.}
  \bibinfo{year}{1996}\natexlab{}.
\newblock \bibinfo{booktitle}{\emph{Coloured Petri Nets - Basic Concepts,
  Analysis Methods and Practical Use}}.
\newblock \bibinfo{publisher}{Springer}.
\newblock


\bibitem[\protect\citeauthoryear{Jiang, Li, Wu, and Zhou}{Jiang
  et~al\mbox{.}}{2018}]%
        {Jiang2018}
\bibfield{author}{\bibinfo{person}{Z Jiang}, \bibinfo{person}{Z Li},
  \bibinfo{person}{N Wu}, {and} \bibinfo{person}{M Zhou}.}
  \bibinfo{year}{2018}\natexlab{}.
\newblock \showarticletitle{{A Petri Net Approach to Fault Diagnosis and
  Restoration for Power Transmission Systems to Avoid the Output Interruption
  of Substations}}.
\newblock \bibinfo{journal}{\emph{IEEE Systems Journal}} \bibinfo{volume}{12},
  \bibinfo{number}{3} (\bibinfo{year}{2018}), \bibinfo{pages}{2566--2576}.
\newblock
\showISSN{1937-9234}
\urldef\tempurl%
\url{https://doi.org/10.1109/JSYST.2017.2682185}
\showDOI{\tempurl}


\bibitem[\protect\citeauthoryear{Kaur, Rana, Kumar, Singh, and Mishra}{Kaur
  et~al\mbox{.}}{2016}]%
        {Kaur2016}
\bibfield{author}{\bibinfo{person}{K Kaur}, \bibinfo{person}{R Rana},
  \bibinfo{person}{N Kumar}, \bibinfo{person}{M Singh}, {and}
  \bibinfo{person}{S Mishra}.} \bibinfo{year}{2016}\natexlab{}.
\newblock \showarticletitle{{A Colored Petri Net Based Frequency Support Scheme
  Using Fleet of Electric Vehicles in Smart Grid Environment}}.
\newblock \bibinfo{journal}{\emph{IEEE Transactions on Power Systems}}
  \bibinfo{volume}{31}, \bibinfo{number}{6} (\bibinfo{year}{2016}),
  \bibinfo{pages}{4638--4649}.
\newblock
\showISSN{1558-0679}
\urldef\tempurl%
\url{https://doi.org/10.1109/TPWRS.2016.2518743}
\showDOI{\tempurl}


\bibitem[\protect\citeauthoryear{Kiaei and Lotfifard}{Kiaei and
  Lotfifard}{2020}]%
        {Kiaei2020}
\bibfield{author}{\bibinfo{person}{I Kiaei} {and} \bibinfo{person}{S
  Lotfifard}.} \bibinfo{year}{2020}\natexlab{}.
\newblock \showarticletitle{{Fault Section Identification in Smart Distribution
  Systems Using Multi-Source Data Based on Fuzzy Petri Nets}}.
\newblock \bibinfo{journal}{\emph{IEEE Transactions on Smart Grid}}
  \bibinfo{volume}{11}, \bibinfo{number}{1} (\bibinfo{year}{2020}),
  \bibinfo{pages}{74--83}.
\newblock
\showISSN{1949-3061}
\urldef\tempurl%
\url{https://doi.org/10.1109/TSG.2019.2917506}
\showDOI{\tempurl}


\bibitem[\protect\citeauthoryear{Kim and Xu}{Kim and Xu}{2019}]%
        {Kim2019}
\bibfield{author}{\bibinfo{person}{Insu Kim} {and} \bibinfo{person}{Shuo Xu}.}
  \bibinfo{year}{2019}\natexlab{}.
\newblock \showarticletitle{{Bus voltage control and optimization strategies
  for power flow analyses using Petri net approach}}.
\newblock \bibinfo{journal}{\emph{International Journal of Electrical Power \&
  Energy Systems}}  \bibinfo{volume}{112} (\bibinfo{year}{2019}),
  \bibinfo{pages}{353--361}.
\newblock
\showISSN{0142-0615}
\urldef\tempurl%
\url{https://doi.org/10.1016/j.ijepes.2019.05.009}
\showDOI{\tempurl}


\bibitem[\protect\citeauthoryear{Kyriakarakos, Dounis, Arvanitis, and
  Papadakis}{Kyriakarakos et~al\mbox{.}}{2012}]%
        {Kyriakarakos2012a}
\bibfield{author}{\bibinfo{person}{George Kyriakarakos},
  \bibinfo{person}{Anastasios~I Dounis}, \bibinfo{person}{Konstantinos~G
  Arvanitis}, {and} \bibinfo{person}{George Papadakis}.}
  \bibinfo{year}{2012}\natexlab{}.
\newblock \showarticletitle{{A fuzzy cognitive maps-petri nets energy
  management system for autonomous polygeneration microgrids}}.
\newblock \bibinfo{journal}{\emph{Appl. Soft Comput.}} \bibinfo{volume}{12},
  \bibinfo{number}{12} (\bibinfo{year}{2012}), \bibinfo{pages}{3785--3797}.
\newblock
\urldef\tempurl%
\url{https://doi.org/10.1016/j.asoc.2012.01.024}
\showDOI{\tempurl}


\bibitem[\protect\citeauthoryear{Lei, Zhang, Shen, Lin, and Zhong}{Lei
  et~al\mbox{.}}{2013}]%
        {Lei13}
\bibfield{author}{\bibinfo{person}{Lei Lei}, \bibinfo{person}{Yingkai Zhang},
  \bibinfo{person}{Xuemin~Sherman Shen}, \bibinfo{person}{Chuang Lin}, {and}
  \bibinfo{person}{Zhangdui Zhong}.} \bibinfo{year}{2013}\natexlab{}.
\newblock \showarticletitle{Performance Analysis of Device-to-Device
  Communications with Dynamic Interference Using Stochastic Petri Nets}.
\newblock \bibinfo{journal}{\emph{IEEE Transactions on Wireless
  Communications}} \bibinfo{volume}{12}, \bibinfo{number}{12}
  (\bibinfo{year}{2013}), \bibinfo{pages}{6121--6141}.
\newblock


\bibitem[\protect\citeauthoryear{Li, Lu, Choo, Wang, and Luo}{Li
  et~al\mbox{.}}{2018}]%
        {li2018reliability}
\bibfield{author}{\bibinfo{person}{Beibei Li}, \bibinfo{person}{Rongxing Lu},
  \bibinfo{person}{Kim-Kwang~Raymond Choo}, \bibinfo{person}{Wei Wang}, {and}
  \bibinfo{person}{Sheng Luo}.} \bibinfo{year}{2018}\natexlab{}.
\newblock \showarticletitle{On reliability analysis of smart grids under
  topology attacks: A stochastic petri net approach}.
\newblock \bibinfo{journal}{\emph{ACM Transactions on Cyber-Physical Systems}}
  \bibinfo{volume}{3}, \bibinfo{number}{1} (\bibinfo{year}{2018}),
  \bibinfo{pages}{1--25}.
\newblock


\bibitem[\protect\citeauthoryear{Li, Lu, Xiao, Bao, and Ghorbani}{Li
  et~al\mbox{.}}{2019}]%
        {Li2019a}
\bibfield{author}{\bibinfo{person}{B Li}, \bibinfo{person}{R Lu},
  \bibinfo{person}{G Xiao}, \bibinfo{person}{H Bao}, {and} \bibinfo{person}{A~A
  Ghorbani}.} \bibinfo{year}{2019}\natexlab{}.
\newblock \showarticletitle{{Towards insider threats detection in smart grid
  communication systems}}.
\newblock \bibinfo{journal}{\emph{IET Communications}} \bibinfo{volume}{13},
  \bibinfo{number}{12} (\bibinfo{year}{2019}), \bibinfo{pages}{1728--1736}.
\newblock
\showISSN{1751-8636}
\urldef\tempurl%
\url{https://doi.org/10.1049/iet-com.2018.5736}
\showDOI{\tempurl}


\bibitem[\protect\citeauthoryear{Liu, Hu, Song, Wang, and Xie}{Liu
  et~al\mbox{.}}{2014}]%
        {liu2014vehicle}
\bibfield{author}{\bibinfo{person}{Hui Liu}, \bibinfo{person}{Zechun Hu},
  \bibinfo{person}{Yonghua Song}, \bibinfo{person}{Jianhui Wang}, {and}
  \bibinfo{person}{Xu Xie}.} \bibinfo{year}{2014}\natexlab{}.
\newblock \showarticletitle{Vehicle-to-grid control for supplementary frequency
  regulation considering charging demands}.
\newblock \bibinfo{journal}{\emph{IEEE Transactions on Power Systems}}
  \bibinfo{volume}{30}, \bibinfo{number}{6} (\bibinfo{year}{2014}),
  \bibinfo{pages}{3110--3119}.
\newblock


\bibitem[\protect\citeauthoryear{Liu, Zhang, and Zhu}{Liu
  et~al\mbox{.}}{2017}]%
        {liu2017modeling}
\bibfield{author}{\bibinfo{person}{Xiaoxue Liu}, \bibinfo{person}{Jiexin
  Zhang}, {and} \bibinfo{person}{Peidong Zhu}.}
  \bibinfo{year}{2017}\natexlab{}.
\newblock \showarticletitle{Modeling cyber-physical attacks based on
  probabilistic colored Petri nets and mixed-strategy game theory}.
\newblock \bibinfo{journal}{\emph{International Journal of Critical
  Infrastructure Protection}}  \bibinfo{volume}{16} (\bibinfo{year}{2017}),
  \bibinfo{pages}{13--25}.
\newblock


\bibitem[\protect\citeauthoryear{Liu, Wang, Peng, and Guo}{Liu
  et~al\mbox{.}}{2010}]%
        {Liu2010}
\bibfield{author}{\bibinfo{person}{Y Liu}, \bibinfo{person}{Y Wang},
  \bibinfo{person}{M Peng}, {and} \bibinfo{person}{C Guo}.}
  \bibinfo{year}{2010}\natexlab{}.
\newblock \showarticletitle{{A fault diagnosis method for power system based on
  multilayer information fusion structure}}. In \bibinfo{booktitle}{\emph{IEEE
  PES General Meeting}}. \bibinfo{publisher}{IEEE}, \bibinfo{pages}{1--5}.
\newblock
\showISSN{1944-9925}
\urldef\tempurl%
\url{https://doi.org/10.1109/PES.2010.5589380}
\showDOI{\tempurl}


\bibitem[\protect\citeauthoryear{Lu, Fakham, Zhou, and Fran{\c{c}}ois}{Lu
  et~al\mbox{.}}{2010}]%
        {Lu2010}
\bibfield{author}{\bibinfo{person}{D Lu}, \bibinfo{person}{H Fakham},
  \bibinfo{person}{T Zhou}, {and} \bibinfo{person}{B Fran{\c{c}}ois}.}
  \bibinfo{year}{2010}\natexlab{}.
\newblock \showarticletitle{{Application of Petri nets for the energy
  management of a photovoltaic based power station including storage units}}.
\newblock \bibinfo{journal}{\emph{Renewable Energy}} \bibinfo{volume}{35},
  \bibinfo{number}{6} (\bibinfo{year}{2010}), \bibinfo{pages}{1117--1124}.
\newblock
\showISSN{0960-1481}
\urldef\tempurl%
\url{https://doi.org/10.1016/j.renene.2009.12.017}
\showDOI{\tempurl}


\bibitem[\protect\citeauthoryear{Machado, Silva, de~Souza, de~Souza, and
  Netto}{Machado et~al\mbox{.}}{2018}]%
        {Machado2018}
\bibfield{author}{\bibinfo{person}{Pedro Machado}, \bibinfo{person}{Milton~R
  Silva}, \bibinfo{person}{Luiz~E de Souza}, \bibinfo{person}{Carlos~W de
  Souza}, {and} \bibinfo{person}{Roberto~S Netto}.}
  \bibinfo{year}{2018}\natexlab{}.
\newblock \showarticletitle{Modeling using colored petri net of communication
  networks based on iec 61850 in a microgrid context}.
\newblock \bibinfo{journal}{\emph{Journal of Control, Automation and Electrical
  Systems}} \bibinfo{volume}{29}, \bibinfo{number}{6} (\bibinfo{year}{2018}),
  \bibinfo{pages}{703--717}.
\newblock


\bibitem[\protect\citeauthoryear{Mahdi, Chalah, and Nadji}{Mahdi
  et~al\mbox{.}}{2017}]%
        {Mahdi2017}
\bibfield{author}{\bibinfo{person}{Ismahan Mahdi}, \bibinfo{person}{Samira
  Chalah}, {and} \bibinfo{person}{Bouchra Nadji}.}
  \bibinfo{year}{2017}\natexlab{}.
\newblock \showarticletitle{{Reliability study of a system dedicated to
  renewable energies by using stochastic petri nets: application to
  photovoltaic (PV) system}}.
\newblock \bibinfo{journal}{\emph{Energy Procedia}}  \bibinfo{volume}{136}
  (\bibinfo{year}{2017}), \bibinfo{pages}{513--520}.
\newblock
\showISSN{1876-6102}
\urldef\tempurl%
\url{https://doi.org/10.1016/j.egypro.2017.10.276}
\showDOI{\tempurl}


\bibitem[\protect\citeauthoryear{Mahendran, Gunasekaran, and Siva
  Ram~Murthy}{Mahendran et~al\mbox{.}}{2014}]%
        {Mahendran14}
\bibfield{author}{\bibinfo{person}{V. Mahendran}, \bibinfo{person}{Rajkishan
  Gunasekaran}, {and} \bibinfo{person}{C. Siva Ram~Murthy}.}
  \bibinfo{year}{2014}\natexlab{}.
\newblock \showarticletitle{Performance Modeling of Delay-Tolerant Network
  Routing via Queueing Petri Nets}.
\newblock \bibinfo{journal}{\emph{IEEE Transactions on Mobile Computing}}
  \bibinfo{volume}{13}, \bibinfo{number}{8} (\bibinfo{year}{2014}),
  \bibinfo{pages}{1816--1828}.
\newblock


\bibitem[\protect\citeauthoryear{Marrone and Gentile}{Marrone and
  Gentile}{2016}]%
        {Marrone2016}
\bibfield{author}{\bibinfo{person}{Stefano Marrone} {and} \bibinfo{person}{Ugo
  Gentile}.} \bibinfo{year}{2016}\natexlab{}.
\newblock \showarticletitle{{Finding Resilient and Energy-saving Control
  Strategies in Smart Homes}}.
\newblock \bibinfo{journal}{\emph{Procedia Computer Science}}
  \bibinfo{volume}{83} (\bibinfo{year}{2016}), \bibinfo{pages}{976--981}.
\newblock
\showISSN{1877-0509}
\urldef\tempurl%
\url{https://doi.org/10.1016/j.procs.2016.04.195}
\showDOI{\tempurl}


\bibitem[\protect\citeauthoryear{Marsan}{Marsan}{1988}]%
        {SPN}
\bibfield{author}{\bibinfo{person}{Marco~Ajmone Marsan}.}
  \bibinfo{year}{1988}\natexlab{}.
\newblock \showarticletitle{Stochastic Petri nets: an elementary introduction}.
  In \bibinfo{booktitle}{\emph{Advances in Petri Nets 1989, covers the 9th
  European Workshop on Applications and Theory in Petri Nets, held in Venice,
  Italy in June 1988, selected papers}} \emph{(\bibinfo{series}{Lecture Notes
  in Computer Science})}, \bibfield{editor}{\bibinfo{person}{Grzegorz
  Rozenberg}} (Ed.), Vol.~\bibinfo{volume}{424}. \bibinfo{publisher}{Springer},
  \bibinfo{pages}{1--29}.
\newblock


\bibitem[\protect\citeauthoryear{Matos and Sanchez}{Matos and Sanchez}{2016}]%
        {Matos2016}
\bibfield{author}{\bibinfo{person}{L~O Matos} {and} \bibinfo{person}{J~W~G
  Sanchez}.} \bibinfo{year}{2016}\natexlab{}.
\newblock \showarticletitle{{Reconfiguration strategy for Fault Tolerance of
  power Distribution Systems using Petri net}}. In
  \bibinfo{booktitle}{\emph{2016 IEEE Ecuador Technical Chapters Meeting
  (ETCM)}}. \bibinfo{pages}{1--6}.
\newblock
\urldef\tempurl%
\url{https://doi.org/10.1109/ETCM.2016.7750820}
\showDOI{\tempurl}


\bibitem[\protect\citeauthoryear{Medina, Aguilar, and
  Rodr{\'{\i}}guez{-}Moreno}{Medina et~al\mbox{.}}{2023}]%
        {MedinaAR23}
\bibfield{author}{\bibinfo{person}{Marcel Sime{\'{o}}n~Garc{\'{\i}}a Medina},
  \bibinfo{person}{Jos{\'{e}} Aguilar}, {and}
  \bibinfo{person}{Mar{\'{\i}}a~Dolores Rodr{\'{\i}}guez{-}Moreno}.}
  \bibinfo{year}{2023}\natexlab{}.
\newblock \showarticletitle{A Bioinspired Emergent Control for Smart Grids}.
\newblock \bibinfo{journal}{\emph{{IEEE} Access}}  \bibinfo{volume}{11}
  (\bibinfo{year}{2023}), \bibinfo{pages}{7503--7520}.
\newblock


\bibitem[\protect\citeauthoryear{Morris, Kim, Wood, and Woodward}{Morris
  et~al\mbox{.}}{2017}]%
        {Morris2017}
\bibfield{author}{\bibinfo{person}{K Morris}, \bibinfo{person}{D~S Kim},
  \bibinfo{person}{A Wood}, {and} \bibinfo{person}{G Woodward}.}
  \bibinfo{year}{2017}\natexlab{}.
\newblock \showarticletitle{{Availability and resiliency analysis of modern
  distribution grids using stochastic reward nets}}. In
  \bibinfo{booktitle}{\emph{2017 IEEE Innovative Smart Grid Technologies - Asia
  (ISGT-Asia)}}. \bibinfo{publisher}{IEEE}, \bibinfo{pages}{1--5}.
\newblock
\showISSN{2378-8542}
\urldef\tempurl%
\url{https://doi.org/10.1109/ISGT-Asia.2017.8378407}
\showDOI{\tempurl}


\bibitem[\protect\citeauthoryear{Munn, Peters, Stern, Tufanaru, McArthur, and
  Aromataris}{Munn et~al\mbox{.}}{2018}]%
        {munn2018systematic}
\bibfield{author}{\bibinfo{person}{Zachary Munn}, \bibinfo{person}{Micah~DJ
  Peters}, \bibinfo{person}{Cindy Stern}, \bibinfo{person}{Catalin Tufanaru},
  \bibinfo{person}{Alexa McArthur}, {and} \bibinfo{person}{Edoardo
  Aromataris}.} \bibinfo{year}{2018}\natexlab{}.
\newblock \showarticletitle{Systematic review or scoping review? Guidance for
  authors when choosing between a systematic or scoping review approach}.
\newblock \bibinfo{journal}{\emph{BMC medical research methodology}}
  \bibinfo{volume}{18} (\bibinfo{year}{2018}), \bibinfo{pages}{1--7}.
\newblock


\bibitem[\protect\citeauthoryear{Mu{\~{n}}oz and {Alvaroy Torres}}{Mu{\~{n}}oz
  and {Alvaroy Torres}}{2010}]%
        {Munoz2010}
\bibfield{author}{\bibinfo{person}{A~D Mu{\~{n}}oz} {and} \bibinfo{person}{M
  {Alvaroy Torres}}.} \bibinfo{year}{2010}\natexlab{}.
\newblock \showarticletitle{{An Intelligent Protection System for a
  transmission network}}. In \bibinfo{booktitle}{\emph{2010 IEEE ANDESCON}}.
  \bibinfo{publisher}{IEEE}, \bibinfo{pages}{1--6}.
\newblock
\urldef\tempurl%
\url{https://doi.org/10.1109/ANDESCON.2010.5633310}
\showDOI{\tempurl}


\bibitem[\protect\citeauthoryear{Panchal and Kumar}{Panchal and Kumar}{2016}]%
        {Panchal2016}
\bibfield{author}{\bibinfo{person}{Dilbagh Panchal} {and}
  \bibinfo{person}{Dinesh Kumar}.} \bibinfo{year}{2016}\natexlab{}.
\newblock \showarticletitle{{Stochastic behaviour analysis of power generating
  unit in thermal power plant using fuzzy methodology}}.
\newblock \bibinfo{journal}{\emph{OPSEARCH}} \bibinfo{volume}{53},
  \bibinfo{number}{1} (\bibinfo{date}{mar} \bibinfo{year}{2016}),
  \bibinfo{pages}{16--40}.
\newblock
\showISSN{09750320}
\urldef\tempurl%
\url{https://doi.org/10.1007/s12597-015-0219-4}
\showDOI{\tempurl}


\bibitem[\protect\citeauthoryear{Pawar and Momin}{Pawar and Momin}{2017}]%
        {ref:pawar2017data-analytics-brief-review}
\bibfield{author}{\bibinfo{person}{Savita Pawar} {and} \bibinfo{person}{BF
  Momin}.} \bibinfo{year}{2017}\natexlab{}.
\newblock \showarticletitle{Smart electricity meter data analytics: A brief
  review}. In \bibinfo{booktitle}{\emph{IEEE Region 10 Symposium (TENSYMP),
  2017}}. IEEE, \bibinfo{publisher}{IEEE}, \bibinfo{pages}{1--5}.
\newblock


\bibitem[\protect\citeauthoryear{Peng, Li, Zhao, Guo, Lv, Tan, and Zhang}{Peng
  et~al\mbox{.}}{2019}]%
        {Peng2019}
\bibfield{author}{\bibinfo{person}{Shitong Peng}, \bibinfo{person}{Tao Li},
  \bibinfo{person}{Jiali Zhao}, \bibinfo{person}{Yanchun Guo},
  \bibinfo{person}{Shengping Lv}, \bibinfo{person}{George~Z. Tan}, {and}
  \bibinfo{person}{Hongchao Zhang}.} \bibinfo{year}{2019}\natexlab{}.
\newblock \showarticletitle{Petri net-based scheduling strategy and energy
  modeling for the cylinder block remanufacturing under uncertainty}.
\newblock \bibinfo{journal}{\emph{Robotics and Computer-Integrated
  Manufacturing}}  \bibinfo{volume}{58} (\bibinfo{year}{2019}),
  \bibinfo{pages}{208--219}.
\newblock


\bibitem[\protect\citeauthoryear{Peters}{Peters}{1999}]%
        {FPN}
\bibfield{author}{\bibinfo{person}{James~F. Peters}.}
  \bibinfo{year}{1999}\natexlab{}.
\newblock \showarticletitle{Introduction: Threads in fuzzy Petri nets
  research}.
\newblock \bibinfo{journal}{\emph{Int. J. Intell. Syst.}} \bibinfo{volume}{14},
  \bibinfo{number}{8} (\bibinfo{year}{1999}), \bibinfo{pages}{717--718}.
\newblock


\bibitem[\protect\citeauthoryear{Petri}{Petri}{1966}]%
        {Ireaditandindeeditisclassic}
\bibfield{author}{\bibinfo{person}{Carl~Adam Petri}.}
  \bibinfo{year}{1966}\natexlab{}.
\newblock \showarticletitle{Communication with automata}.
\newblock


\bibitem[\protect\citeauthoryear{Qachchachi, Mahmoudi, and Hasnaoui}{Qachchachi
  et~al\mbox{.}}{2016}]%
        {Qachchachi2016}
\bibfield{author}{\bibinfo{person}{N Qachchachi}, \bibinfo{person}{H Mahmoudi},
  {and} \bibinfo{person}{A~E Hasnaoui}.} \bibinfo{year}{2016}\natexlab{}.
\newblock \showarticletitle{{Smart hybrid AC/DC microgrid: Power management
  based Petri Nets}}. In \bibinfo{booktitle}{\emph{2016 International
  Conference on Information Technology for Organizations Development (IT4OD)}}.
  \bibinfo{pages}{1--6}.
\newblock
\urldef\tempurl%
\url{https://doi.org/10.1109/IT4OD.2016.7479320}
\showDOI{\tempurl}


\bibitem[\protect\citeauthoryear{{Rafael da Silva}, {Ferreira Machado},
  de~Souza, and de~Souza}{{Rafael da Silva} et~al\mbox{.}}{2017}]%
        {RafaeldaSilva2017}
\bibfield{author}{\bibinfo{person}{M {Rafael da Silva}}, \bibinfo{person}{P~H
  {Ferreira Machado}}, \bibinfo{person}{L~E de Souza}, {and}
  \bibinfo{person}{C~W de Souza}.} \bibinfo{year}{2017}\natexlab{}.
\newblock \showarticletitle{{Modeling of a cyber-attack in an IEC 61850
  scenario using stochastic colored Petri Nets}}. In
  \bibinfo{booktitle}{\emph{2017 4th International Conference on Systems and
  Informatics (ICSAI)}}. \bibinfo{pages}{985--990}.
\newblock
\urldef\tempurl%
\url{https://doi.org/10.1109/ICSAI.2017.8248429}
\showDOI{\tempurl}


\bibitem[\protect\citeauthoryear{Reisig}{Reisig}{2013}]%
        {mouzhianotheronereadinprogress}
\bibfield{author}{\bibinfo{person}{Wolfgang Reisig}.}
  \bibinfo{year}{2013}\natexlab{}.
\newblock \bibinfo{booktitle}{\emph{Understanding Petri Nets - Modeling
  Techniques, Analysis Methods, Case Studies}}.
\newblock \bibinfo{publisher}{Springer}.
\newblock


\bibitem[\protect\citeauthoryear{{Rossi} and {Chren}}{{Rossi} and
  {Chren}}{2020}]%
        {ref:rossi2020}
\bibfield{author}{\bibinfo{person}{B. {Rossi}} {and} \bibinfo{person}{S.
  {Chren}}.} \bibinfo{year}{2020}\natexlab{}.
\newblock \showarticletitle{Smart Grids Data Analysis: A Systematic Mapping
  Study}.
\newblock \bibinfo{journal}{\emph{IEEE Transactions on Industrial Informatics}}
  \bibinfo{volume}{16}, \bibinfo{number}{6} (\bibinfo{year}{2020}),
  \bibinfo{pages}{3619--3639}.
\newblock
\urldef\tempurl%
\url{https://doi.org/10.1109/TII.2019.2954098}
\showDOI{\tempurl}


\bibitem[\protect\citeauthoryear{Rozenberg and Engelfriet}{Rozenberg and
  Engelfriet}{1998}]%
        {Rozenberg1998}
\bibfield{author}{\bibinfo{person}{Grzegorz Rozenberg} {and}
  \bibinfo{person}{Joost Engelfriet}.} \bibinfo{year}{1998}\natexlab{}.
\newblock \bibinfo{booktitle}{\emph{Elementary net systems}}.
\newblock \bibinfo{publisher}{Springer Berlin Heidelberg},
  \bibinfo{address}{Berlin, Heidelberg}, \bibinfo{pages}{12--121}.
\newblock


\bibitem[\protect\citeauthoryear{Saki, Fereidunian, Lesani, and
  Kolarijani}{Saki et~al\mbox{.}}{2011}]%
        {Saki2011}
\bibfield{author}{\bibinfo{person}{M Saki}, \bibinfo{person}{A Fereidunian},
  \bibinfo{person}{H Lesani}, {and} \bibinfo{person}{M~A~S Kolarijani}.}
  \bibinfo{year}{2011}\natexlab{}.
\newblock \showarticletitle{{Distribution automation monitoring using Petri
  nets}}. In \bibinfo{booktitle}{\emph{The 2nd International Conference on
  Control, Instrumentation and Automation}}. \bibinfo{pages}{56--61}.
\newblock
\urldef\tempurl%
\url{https://doi.org/10.1109/ICCIAutom.2011.6356630}
\showDOI{\tempurl}


\bibitem[\protect\citeauthoryear{Shateri, Messina, Piantanida, and
  Labeau}{Shateri et~al\mbox{.}}{2020}]%
        {ShateriMPL20}
\bibfield{author}{\bibinfo{person}{Mohammadhadi Shateri},
  \bibinfo{person}{Francisco Messina}, \bibinfo{person}{Pablo Piantanida},
  {and} \bibinfo{person}{Fabrice Labeau}.} \bibinfo{year}{2020}\natexlab{}.
\newblock \showarticletitle{Real-Time Privacy-Preserving Data Release for Smart
  Meters}.
\newblock \bibinfo{journal}{\emph{{IEEE} Trans. Smart Grid}}
  \bibinfo{volume}{11}, \bibinfo{number}{6} (\bibinfo{year}{2020}),
  \bibinfo{pages}{5174--5183}.
\newblock


\bibitem[\protect\citeauthoryear{Springmann, Bruckmeier, and
  M{\"{u}}ller}{Springmann et~al\mbox{.}}{2022}]%
        {SpringmannBM22}
\bibfield{author}{\bibinfo{person}{Elisabeth Springmann},
  \bibinfo{person}{Andreas Bruckmeier}, {and} \bibinfo{person}{Mathias
  M{\"{u}}ller}.} \bibinfo{year}{2022}\natexlab{}.
\newblock \showarticletitle{Performance evaluation of German smart meter
  infrastructure for load management through grid operators}.
\newblock \bibinfo{journal}{\emph{Energy Inform.}}  \bibinfo{volume}{5}
  (\bibinfo{year}{2022}).
\newblock


\bibitem[\protect\citeauthoryear{Sreerama and Swarup}{Sreerama and
  Swarup}{2017}]%
        {Sreerama2017}
\bibfield{author}{\bibinfo{person}{R Sreerama} {and} \bibinfo{person}{K~S
  Swarup}.} \bibinfo{year}{2017}\natexlab{}.
\newblock \showarticletitle{{Detection, localization and fault diagnosis using
  PetriNets for smart power distribution grids}}. In
  \bibinfo{booktitle}{\emph{2017 7th International Conference on Power Systems
  (ICPS)}}. \bibinfo{pages}{596--600}.
\newblock
\urldef\tempurl%
\url{https://doi.org/10.1109/ICPES.2017.8387363}
\showDOI{\tempurl}


\bibitem[\protect\citeauthoryear{Suhaimy, Radzi, Ahmad, Azmi, and
  Hannan}{Suhaimy et~al\mbox{.}}{2022}]%
        {SuhaimyRAAH22}
\bibfield{author}{\bibinfo{person}{Nurshazlina Suhaimy}, \bibinfo{person}{Nurul
  Asyikin~Mohamed Radzi}, \bibinfo{person}{Wan Siti Halimatul Munirah~Wan
  Ahmad}, \bibinfo{person}{Kaiyisah Hanis~Mohd Azmi}, {and}
  \bibinfo{person}{M.~A. Hannan}.} \bibinfo{year}{2022}\natexlab{}.
\newblock \showarticletitle{Current and Future Communication Solutions for
  Smart Grids: {A} Review}.
\newblock \bibinfo{journal}{\emph{{IEEE} Access}}  \bibinfo{volume}{10}
  (\bibinfo{year}{2022}), \bibinfo{pages}{43639--43668}.
\newblock


\bibitem[\protect\citeauthoryear{Tare, Waghmare, Siddavatam, Kazi, and
  Singh}{Tare et~al\mbox{.}}{2016}]%
        {Tare2016}
\bibfield{author}{\bibinfo{person}{B Tare}, \bibinfo{person}{S Waghmare},
  \bibinfo{person}{I Siddavatam}, \bibinfo{person}{F Kazi}, {and}
  \bibinfo{person}{N Singh}.} \bibinfo{year}{2016}\natexlab{}.
\newblock \showarticletitle{{Security analysis of DNP3 using CPN model with
  state space report representation using LDA}}. In
  \bibinfo{booktitle}{\emph{2016 Indian Control Conference (ICC)}}.
  \bibinfo{pages}{25--31}.
\newblock
\urldef\tempurl%
\url{https://doi.org/10.1109/INDIANCC.2016.7441101}
\showDOI{\tempurl}


\bibitem[\protect\citeauthoryear{Tolosana-Calasanz, Ban{\~{a}}res, Cipcigan,
  Rana, Papadopoulos, and Pham}{Tolosana-Calasanz et~al\mbox{.}}{2013}]%
        {Tolosana-Calasanz2013}
\bibfield{author}{\bibinfo{person}{Rafael Tolosana-Calasanz},
  \bibinfo{person}{Jos{\'{e}}~Angel Ban{\~{a}}res}, \bibinfo{person}{Liana
  Cipcigan}, \bibinfo{person}{Omer Rana}, \bibinfo{person}{Panagiotis
  Papadopoulos}, {and} \bibinfo{person}{Congduc Pham}.}
  \bibinfo{year}{2013}\natexlab{}.
\newblock \showarticletitle{{A distributed in-transit processing infrastructure
  for forecasting electric vehicle charging demand}}.
\newblock \bibinfo{journal}{\emph{Proceedings - 13th IEEE/ACM International
  Symposium on Cluster, Cloud, and Grid Computing, CCGrid 2013}}
  (\bibinfo{year}{2013}), \bibinfo{pages}{538--545}.
\newblock
\showISBNx{9780769549965}
\urldef\tempurl%
\url{https://doi.org/10.1109/CCGrid.2013.103}
\showDOI{\tempurl}


\bibitem[\protect\citeauthoryear{Tu, He, Shuai, and Jiang}{Tu
  et~al\mbox{.}}{2017}]%
        {ref:tu2017bigdatareview}
\bibfield{author}{\bibinfo{person}{Chunming Tu}, \bibinfo{person}{Xi He},
  \bibinfo{person}{Zhikang Shuai}, {and} \bibinfo{person}{Fei Jiang}.}
  \bibinfo{year}{2017}\natexlab{}.
\newblock \showarticletitle{Big data issues in smart grid--A review}.
\newblock \bibinfo{journal}{\emph{Renewable and Sustainable Energy Reviews}}
  \bibinfo{volume}{79} (\bibinfo{year}{2017}), \bibinfo{pages}{1099--1107}.
\newblock


\bibitem[\protect\citeauthoryear{Wang}{Wang}{1998}]%
        {TPN}
\bibfield{author}{\bibinfo{person}{Jiacun Wang}.}
  \bibinfo{year}{1998}\natexlab{}.
\newblock \bibinfo{booktitle}{\emph{Timed Petri NetS - Theory and
  Application}}.
\newblock \bibinfo{publisher}{Springer}.
\newblock


\bibitem[\protect\citeauthoryear{Wang, Meng, Cao, Chen, Gao, and Lin}{Wang
  et~al\mbox{.}}{2014a}]%
        {Wang2014a}
\bibfield{author}{\bibinfo{person}{Jiye Wang}, \bibinfo{person}{Kun Meng},
  \bibinfo{person}{Junwei Cao}, \bibinfo{person}{Zhen Chen},
  \bibinfo{person}{Lingchao Gao}, {and} \bibinfo{person}{Chuang Lin}.}
  \bibinfo{year}{2014}\natexlab{a}.
\newblock \showarticletitle{{Electricity Services Based Dependability Model of
  Power Grid Communication Networking}}.
\newblock \bibinfo{journal}{\emph{TSINGHUA SCIENCE AND TECHNOLOGY}}
  \bibinfo{volume}{19}, \bibinfo{number}{2} (\bibinfo{date}{apr}
  \bibinfo{year}{2014}), \bibinfo{pages}{121--132}.
\newblock
\showISSN{1007-0214}


\bibitem[\protect\citeauthoryear{Wang, Chen, Gao, Niu, Zhao, Ma, and Wu}{Wang
  et~al\mbox{.}}{2015}]%
        {wang2015a}
\bibfield{author}{\bibinfo{person}{L Wang}, \bibinfo{person}{Q Chen},
  \bibinfo{person}{Z Gao}, \bibinfo{person}{L Niu}, \bibinfo{person}{Y Zhao},
  \bibinfo{person}{Z Ma}, {and} \bibinfo{person}{D Wu}.}
  \bibinfo{year}{2015}\natexlab{}.
\newblock \showarticletitle{{Knowledge representation and general Petri net
  models for power grid fault diagnosis}}.
\newblock \bibinfo{journal}{\emph{IET Generation, Transmission Distribution}}
  \bibinfo{volume}{9}, \bibinfo{number}{9} (\bibinfo{year}{2015}),
  \bibinfo{pages}{866--873}.
\newblock
\showISSN{1751-8695}
\urldef\tempurl%
\url{https://doi.org/10.1049/iet-gtd.2014.0659}
\showDOI{\tempurl}


\bibitem[\protect\citeauthoryear{Wang, Ye, Xu, Chen, Li, and Liu}{Wang
  et~al\mbox{.}}{2014b}]%
        {Wang2014}
\bibfield{author}{\bibinfo{person}{Y~N Wang}, \bibinfo{person}{J~F Ye},
  \bibinfo{person}{G~J Xu}, \bibinfo{person}{Q~M Chen}, \bibinfo{person}{H~Y
  Li}, {and} \bibinfo{person}{X~R Liu}.} \bibinfo{year}{2014}\natexlab{b}.
\newblock \showarticletitle{{Novel hierarchical fault diagnosis approach for
  smart power grid with information fusion of multi-data resources based on
  fuzzy petri net}}. In \bibinfo{booktitle}{\emph{2014 IEEE International
  Conference on Fuzzy Systems (FUZZ-IEEE)}}. \bibinfo{pages}{1183--1189}.
\newblock
\showISSN{1098-7584}
\urldef\tempurl%
\url{https://doi.org/10.1109/FUZZ-IEEE.2014.6891791}
\showDOI{\tempurl}


\bibitem[\protect\citeauthoryear{Xiang, Tauch, and Liu}{Xiang
  et~al\mbox{.}}{2014}]%
        {Xiang2014}
\bibfield{author}{\bibinfo{person}{M Xiang}, \bibinfo{person}{S Tauch}, {and}
  \bibinfo{person}{W Liu}.} \bibinfo{year}{2014}\natexlab{}.
\newblock \showarticletitle{{Dependability and Resource Optimation Analysis for
  Smart Grid Communication Networks}}. In \bibinfo{booktitle}{\emph{2014 IEEE
  Fourth International Conference on Big Data and Cloud Computing}}.
  \bibinfo{pages}{676--681}.
\newblock
\urldef\tempurl%
\url{https://doi.org/10.1109/BDCloud.2014.115}
\showDOI{\tempurl}


\bibitem[\protect\citeauthoryear{Xu, Yin, Yin, Wang, and Pang}{Xu
  et~al\mbox{.}}{2019}]%
        {Xu2019}
\bibfield{author}{\bibinfo{person}{B Xu}, \bibinfo{person}{X Yin},
  \bibinfo{person}{X Yin}, \bibinfo{person}{Y Wang}, {and} \bibinfo{person}{S
  Pang}.} \bibinfo{year}{2019}\natexlab{}.
\newblock \showarticletitle{{Fault Diagnosis of Power Systems Based on Temporal
  Constrained Fuzzy Petri Nets}}.
\newblock \bibinfo{journal}{\emph{IEEE Access}}  \bibinfo{volume}{7}
  (\bibinfo{year}{2019}), \bibinfo{pages}{101895--101904}.
\newblock
\showISSN{2169-3536}
\urldef\tempurl%
\url{https://doi.org/10.1109/ACCESS.2019.2930545}
\showDOI{\tempurl}


\bibitem[\protect\citeauthoryear{Xu and Fu}{Xu and Fu}{2018}]%
        {xu2018petri}
\bibfield{author}{\bibinfo{person}{Yue Xu} {and} \bibinfo{person}{Rong Fu}.}
  \bibinfo{year}{2018}\natexlab{}.
\newblock \showarticletitle{Petri net-based power CPS network attack and impact
  modeling}. In \bibinfo{booktitle}{\emph{2018 5th IEEE International
  Conference on Cloud Computing and Intelligence Systems (CCIS)}}. IEEE,
  \bibinfo{pages}{1107--1110}.
\newblock


\bibitem[\protect\citeauthoryear{Yan, Qian, Sharif, and Tipper}{Yan
  et~al\mbox{.}}{2013}]%
        {ref:yan2013communicationsurvey}
\bibfield{author}{\bibinfo{person}{Ye Yan}, \bibinfo{person}{Yi Qian},
  \bibinfo{person}{Hamid Sharif}, {and} \bibinfo{person}{David Tipper}.}
  \bibinfo{year}{2013}\natexlab{}.
\newblock \showarticletitle{A survey on smart grid communication
  infrastructures: Motivations, requirements and challenges}.
\newblock \bibinfo{journal}{\emph{IEEE communications surveys \& tutorials}}
  \bibinfo{volume}{15}, \bibinfo{number}{1} (\bibinfo{year}{2013}),
  \bibinfo{pages}{5--20}.
\newblock


\bibitem[\protect\citeauthoryear{Yu, Cecati, Dillon, and Simoes}{Yu
  et~al\mbox{.}}{2011}]%
        {yu2011new}
\bibfield{author}{\bibinfo{person}{Xinghuo Yu}, \bibinfo{person}{Carlo Cecati},
  \bibinfo{person}{Tharam Dillon}, {and} \bibinfo{person}{M~Godoy Simoes}.}
  \bibinfo{year}{2011}\natexlab{}.
\newblock \showarticletitle{The new frontier of smart grids}.
\newblock \bibinfo{journal}{\emph{IEEE Industrial Electronics Magazine}}
  \bibinfo{volume}{5}, \bibinfo{number}{3} (\bibinfo{year}{2011}),
  \bibinfo{pages}{49--63}.
\newblock


\bibitem[\protect\citeauthoryear{Zaitsev, Shmeleva, Retschitzegger, and
  Pr{\"{o}}ll}{Zaitsev et~al\mbox{.}}{2016}]%
        {zaitsev2016}
\bibfield{author}{\bibinfo{person}{D.~A. Zaitsev}, \bibinfo{person}{T.~R.
  Shmeleva}, \bibinfo{person}{W. Retschitzegger}, {and} \bibinfo{person}{B.
  Pr{\"{o}}ll}.} \bibinfo{year}{2016}\natexlab{}.
\newblock \showarticletitle{{Security of grid structures under disguised
  traffic attacks}}.
\newblock \bibinfo{journal}{\emph{Cluster Computing}} \bibinfo{volume}{19},
  \bibinfo{number}{3} (\bibinfo{date}{sep} \bibinfo{year}{2016}),
  \bibinfo{pages}{1183--1200}.
\newblock
\showISSN{15737543}
\urldef\tempurl%
\url{https://doi.org/10.1007/s10586-016-0582-9}
\showDOI{\tempurl}


\bibitem[\protect\citeauthoryear{Zamani, Fereidunian, Mansouri, {Sharifi K},
  Boroomand, and Lesani}{Zamani et~al\mbox{.}}{2011}]%
        {Zamani2011}
\bibfield{author}{\bibinfo{person}{M~A Zamani}, \bibinfo{person}{A
  Fereidunian}, \bibinfo{person}{S~S Mansouri}, \bibinfo{person}{M~A {Sharifi
  K}}, \bibinfo{person}{F Boroomand}, {and} \bibinfo{person}{H Lesani}.}
  \bibinfo{year}{2011}\natexlab{}.
\newblock \showarticletitle{{A Petri Net-T3{S}D policy driven method for IT
  infrastructure selection in smart grid}}. In \bibinfo{booktitle}{\emph{2011
  IEEE International Systems Conference, SysCon 2011 - Proceedings}}.
  \bibinfo{pages}{520--526}.
\newblock
\urldef\tempurl%
\url{https://doi.org/10.1109/SYSCON.2011.5929112}
\showDOI{\tempurl}


\bibitem[\protect\citeauthoryear{Zamani, Fereidunian, {Sharifi K.}, Lesani, and
  Lucas}{Zamani et~al\mbox{.}}{2010}]%
        {Zamani2010}
\bibfield{author}{\bibinfo{person}{M~A Zamani}, \bibinfo{person}{A
  Fereidunian}, \bibinfo{person}{M~A {Sharifi K.}}, \bibinfo{person}{H Lesani},
  {and} \bibinfo{person}{C Lucas}.} \bibinfo{year}{2010}\natexlab{}.
\newblock \showarticletitle{{AAPNES: A Petri Net expert system realization of
  adaptive autonomy in smart grid}}. In \bibinfo{booktitle}{\emph{2010 5th
  International Symposium on Telecommunications}}. \bibinfo{pages}{968--973}.
\newblock
\urldef\tempurl%
\url{https://doi.org/10.1109/ISTEL.2010.5734162}
\showDOI{\tempurl}


\bibitem[\protect\citeauthoryear{Zeineb, Sajeh, and Belhassen}{Zeineb
  et~al\mbox{.}}{2016}]%
        {Zeineb2016}
\bibfield{author}{\bibinfo{person}{Mhadhbi Zeineb}, \bibinfo{person}{Zairi
  Sajeh}, {and} \bibinfo{person}{Zouari Belhassen}.}
  \bibinfo{year}{2016}\natexlab{}.
\newblock \showarticletitle{{Generic colored petri nets modeling approach for
  performance analysis of smart grid system}}.
\newblock \bibinfo{journal}{\emph{IREC 2016 - 7th International Renewable
  Energy Congress}} (\bibinfo{year}{2016}).
\newblock
\showISBNx{9781467397674}
\urldef\tempurl%
\url{https://doi.org/10.1109/IREC.2016.7478905}
\showDOI{\tempurl}


\bibitem[\protect\citeauthoryear{Zeng, Jiang, Lin, and Shen}{Zeng
  et~al\mbox{.}}{2011}]%
        {Zeng2011}
\bibfield{author}{\bibinfo{person}{Rongfei Zeng}, \bibinfo{person}{Yixin
  Jiang}, \bibinfo{person}{Chuang Lin}, {and} \bibinfo{person}{Xuemin Shen}.}
  \bibinfo{year}{2011}\natexlab{}.
\newblock \showarticletitle{{A stochastic Petri nets approach to dependability
  analysis of control center networks in smart grid}}.
\newblock \bibinfo{journal}{\emph{2011 International Conference on Wireless
  Communications and Signal Processing, WCSP 2011}} (\bibinfo{year}{2011}),
  \bibinfo{pages}{1--5}.
\newblock
\showISBNx{9781457710100}
\urldef\tempurl%
\url{https://doi.org/10.1109/WCSP.2011.6096966}
\showDOI{\tempurl}


\bibitem[\protect\citeauthoryear{Zeng, Jiang, Lin, and Shen}{Zeng
  et~al\mbox{.}}{2012}]%
        {Zeng2012}
\bibfield{author}{\bibinfo{person}{Rongfei Zeng}, \bibinfo{person}{Yixin
  Jiang}, \bibinfo{person}{Chuang Lin}, {and} \bibinfo{person}{Xuemin~(Sherman)
  Shen}.} \bibinfo{year}{2012}\natexlab{}.
\newblock \showarticletitle{{Dependability Analysis of Control Center Networks
  in Smart Grid Using Stochastic Petri Nets}}.
\newblock \bibinfo{journal}{\emph{IEEE Transactions on Parallel and Distributed
  Systems}} \bibinfo{volume}{23}, \bibinfo{number}{9} (\bibinfo{year}{2012}),
  \bibinfo{pages}{1721--1730}.
\newblock
\showISBNx{9781457710100}
\showISSN{1558-2183}
\urldef\tempurl%
\url{https://doi.org/10.1109/TPDS.2012.68}
\showDOI{\tempurl}


\bibitem[\protect\citeauthoryear{Zhang, Wang, Du, Qian, and Yang}{Zhang
  et~al\mbox{.}}{2018}]%
        {Zhang2018a}
\bibfield{author}{\bibinfo{person}{Yingfeng Zhang}, \bibinfo{person}{Wenbo
  Wang}, \bibinfo{person}{Wei Du}, \bibinfo{person}{Cheng Qian}, {and}
  \bibinfo{person}{Haidong Yang}.} \bibinfo{year}{2018}\natexlab{}.
\newblock \showarticletitle{{Coloured Petri net-based active sensing system of
  real-time and multi-source manufacturing information for smart factory}}.
\newblock \bibinfo{journal}{\emph{International Journal of Advanced
  Manufacturing Technology}} \bibinfo{volume}{94}, \bibinfo{number}{9-12}
  (\bibinfo{date}{feb} \bibinfo{year}{2018}), \bibinfo{pages}{3427--3439}.
\newblock
\showISSN{14333015}


\end{thebibliography}
